\documentstyle[preprint,prd,aps]{revtex}
\input psbox

\begin{document}

\tighten
\title{
\begin{flushright}
CLNS 94/1285  \\
CLEO 94-27 \\
\end{flushright}
Measurement of the $\bar{B} \rightarrow D^*~\ell~\bar{\nu}$
Branching Fractions and $|V_{cb}|$}
\author{
B.~Barish,$^{1}$ M.~Chadha,$^{1}$ S.~Chan,$^{1}$ D.F.~Cowen,$^{1}$
G.~Eigen,$^{1}$ J.S.~Miller,$^{1}$ C.~O'Grady,$^{1}$ J.~Urheim,$^{1}$
A.J.~Weinstein,$^{1}$
D.~Acosta,$^{2}$ M.~Athanas,$^{2}$
% ...
}

\address{
\bigskip 
{\rm (CLEO Collaboration)}\\  
\newpage 
$^{1}${California Institute of Technology, Pasadena, California 91125}\\
$^{2}${University of California, San Diego, La Jolla, California 92093}\\
% ...
\bigskip 
}        
\date{\today}
\maketitle
\begin{abstract}
We study the exclusive semileptonic
$B$ meson decays $B^- \rightarrow D^{*0}~\ell^-~\bar{\nu}$ and
$\bar{B}^0 \rightarrow D^{*+}~\ell^-~\bar{\nu}$ using data collected
with the CLEO II detector at the Cornell Electron-positron Storage
Ring (CESR). We present measurements of the branching fractions,
${\cal B}(\bar{B}^0 \rightarrow D^{*+}~\ell^-~\bar{\nu}) =
(0.5/f_{00})[4.49 \pm 0.32({\rm stat.}) \pm 0.39({\rm sys.})]\%$ and
${\cal B}(B^- \rightarrow D^{*0}~\ell^-~\bar{\nu}) =
(0.5/f_{+-})[5.13 \pm 0.54({\rm stat.}) \pm 0.64({\rm sys.})]\%$,
where $f_{00}$ and
$f_{+-}$ are the neutral and charged $B$ meson production
fractions at the $\Upsilon(4S)$ resonance, respectively.
Assuming isospin invariance and taking the
ratio of charged to neutral $B$ meson lifetimes measured
at higher energy machines, we determine the ratio
$f_{+-}/f_{00} = 1.04 \pm 0.13{\rm(stat.)}
                      \pm0.12{\rm (sys.)}
                      \pm0.10{\rm (lifetime)}$;
further assuming $f_{+-}+f_{00}=1$ we also determine
the partial width
$\Gamma(\bar{B} \rightarrow D^*~\ell~\bar{\nu})=
[29.9 \pm 1.9({\rm stat.}) \pm2.7({\rm sys.})
\pm2.0({\rm lifetime})]$~ns$^{-1}$ (independent of
$f_{+-}/f_{00}$). From this partial width we calculate
$\bar{B} \rightarrow D^*~\ell~\bar{\nu}$
branching fractions that do not depend on $f_{+-}/f_{00}$
nor the individual $B$ lifetimes, but
only on the charged to neutral $B$ lifetime ratio.
The product of the CKM matrix element $|V_{cb}|$ times the
normalization of the decay form factor at the point of no recoil of the
$D^*$ meson, ${\cal F}(y=1)$, is determined
from a linear fit to the combined differential decay
rate of the exclusive $\bar{B} \rightarrow D^*~\ell~\bar{\nu}$ decays:
$|V_{cb}|{\cal F}(1)= 0.0351 \pm 0.0019{\rm (stat.)}
\pm 0.0018{\rm (sys.)} \pm 0.0008{\rm (lifetime)}$.
Using theoretical calculations of the form factor normalization we extract
a value for $|V_{cb}|$.
\end{abstract}

\section{introduction}

In the framework of the Standard Model of weak interactions the
elements of the $3~ \times~ 3$ Cabibbo-Kobayashi-Maskawa (CKM) mixing
matrix~\cite{ckm} must be determined experimentally.
The element $|V_{cb}|$ is determined from studies of the  semileptonic decays
of $B$ mesons.  Measurements of $|V_{cb}|$ from the
inclusive semileptonic rate~\cite{arginc,clincold,clinc}
and from exclusive
rates~\cite{oldargusI,oldcleo,crawford,oldargusII,oldargusIII,arguspartial}
are systematically limited by model dependence in the
theoretical prediction of the decay rate.
The recent development of Heavy Quark Effective Theory (HQET)~\cite{wisger}
yields an expression for the
$\bar{B} \rightarrow D^*~\ell~\bar{\nu}$~\cite{ell} decay rate in terms of a
single unknown form factor~\cite{moreneubert} which,
at the point of no recoil of the
$D^*$ meson, is absolutely normalized
up to corrections of order $1/m_Q^2$~\cite{luke}
(where $m_Q$ is the $b$ or $c$ quark mass).
It is currently believed that these corrections
can be calculated with less than
$5\%$ uncertainty~\cite{ffcorr,mannel}, which would permit a
precise determination of $|V_{cb}|$ from the study of
$\bar{B} \rightarrow D^*~\ell~\bar{\nu}$ as a function of
the recoil of the $D^*$ meson.  The decay mode
$\bar{B} \rightarrow D^*~\ell~\bar{\nu}$ is also preferred over other
exclusive channels because $D^*$ meson decays have a
very clean experimental signature.

Throughout this paper the square of the four momentum
transfer in  $\bar{B} \rightarrow D^*~\ell~\bar{\nu}$ decays is
denoted by $q^2 = M^2_{l\bar{\nu}}$
(where $M_{l\bar{\nu}}$ is the mass of the virtual $W$). The kinematic
variable of HQET, which is a measure of the recoil of the
$D^*$ meson, is given by
\begin{equation}
y \equiv v \cdot v' = { (m_B^2 + m_{D^*}^2 - q^2) \over (2m_B m_{D^*})},
\label{eqn:three} \end{equation}
where $v$ and $v'$ are the four velocities of the
$B$ and $D^*$ mesons and $m_B$ and $m_{D^*}$ are their respective masses.

We report on new measurements of the branching fractions and differential
decay rate for the decays,
$\bar{B}^0 \rightarrow D^{*+}~\ell^-~\bar{\nu}$ and
$B^- \rightarrow D^{*0}~\ell^-~\bar{\nu}$~\cite{chgconj}.
The CKM matrix element $|V_{cb}|$ is
extracted from fits to the differential
decay rates as well as the integrated rate.
Using isospin invariance to equate the partial widths of
$\bar{B}^0 \rightarrow D^{*+}~\ell^-~\bar{\nu}$ and
$B^- \rightarrow D^{*0}~\ell^-~\bar{\nu}$, and recent
$B$ meson lifetime measurements,
we also measure the $\Upsilon(4S)$ branching fractions,
\begin{eqnarray}
f_{00} &\equiv& {\cal B}(\Upsilon(4S) \to B^0 {\bar B}^0) \\
f_{+-} &\equiv& {\cal B}(\Upsilon(4S) \to B^- B^+).
\end{eqnarray}
All measurements of
exclusive $B$ branching fractions at the $\Upsilon(4S)$
resonance currently assume equal production of charged and
neutral $B$ mesons, so the
measurement of these production fractions affects  both
hadronic and semileptonic $B$ branching fractions.
A model dependent result for the inclusive branching fraction
${\cal B}(\bar{B} \rightarrow D^*~X~\ell~\bar{\nu})$, where $X$ is
any possible hadronic state ($X \neq 0$),
is also presented.

The paper is structured in the following way: in Section II
the technique for obtaining yields of
$\bar{B}^0 \rightarrow D^{*+}~\ell^-~\bar{\nu}$ and
$B^- \rightarrow D^{*0}~\ell^-~\bar{\nu}$
events is discussed in general.  Details of the candidate selection
are then given in Section III. Backgrounds
in the $\bar{B} \rightarrow D^*~\ell~\bar{\nu}$ samples and how
their magnitudes are estimated
are discussed in Section IV. Systematic studies of the reconstruction
efficiencies are described in section V.
Branching fraction results are presented in Section VI, followed by
measurements of $|V_{cb}|$ in Section VII, and conclusions in Section VIII.


\section{method}


The $\bar{B}^0 \rightarrow D^{*+}~\ell^-~\bar{\nu}$ and
$B^- \rightarrow D^{*0}~\ell^-~\bar{\nu}$ efficiency-corrected yields,
denoted by
$N_0$ and $N_-$ respectively, depend on the total number of $\Upsilon(4S)$
decays in the data sample, $N_{\Upsilon(4S)}$, and a product of
branching fractions,
\begin{eqnarray}
N_0 & = & 4
N_{\Upsilon(4S)}  f_{00}
{\cal B}(\bar{B}^0 \rightarrow D^{*+}~\ell^-~\bar{\nu})
 {\cal B}_{D^{*+}} {\cal B}_{D^0}   \label{eqn:nzero} \\
N_- & = & 4
N_{\Upsilon(4S)}  f_{+-}  {\cal B}(B^- \rightarrow D^{*0}~\ell^-~\bar{\nu})
{\cal B}_{D^{*0}} {\cal B}_{D^0}, \label{eqn:nminus}
\end{eqnarray}
where
$ {\cal B}_{D^{*+}} \equiv {\cal B}(D^{*+} \to D^0\pi^+)$,
$ {\cal B}_{D^{*0}} \equiv {\cal B}(D^{*0} \to D^0\pi^0)$ and
$ {\cal B}_{D^0} \equiv {\cal B}(D^0 \to K^-\pi^+)$. These three branching
fractions have been
measured by CLEO II \cite{dstar,kpi} with $D^*$ samples that are statistically
independent of the sample considered here.
The factors of $4$ enter because each $\Upsilon(4S)$ decay produces
a $B{\bar{B}}$
meson pair and because we are combining the  $e$ and $\mu$
lepton species.

We search for the decays $\bar{B} \rightarrow D^*~\ell~\bar{\nu}$ by combining
reconstructed $D^*$ mesons with ``right sign'' lepton candidates
in the same event.
By ``right sign'' we mean that a $D^{*+}$ ($D^{*0}$) must be paired
with a negative lepton, while a $D^{*-}$ ($\bar{D}^{*0}$) requires a
positive lepton. $D^*$ mesons are reconstructed using the decay chains
$D^{*+} \to D^0 \pi^+$, $D^{*0} \to D^0 \pi^0$ and $D^0 \to K^-\pi^+$
\cite{othermodes}.
This technique uses our knowledge of the B meson
momentum, $|{\bf p}_B|$. The energy of $B$ mesons produced in symmetric
$e^+e^-$
annihilations, $E_B$, must equal the beam energy, which
is precisely known from machine optics;  hence $|{\bf p}_B|$ can be determined
from $E_B$ and the known $B$ mass \cite{bigb}.
A kinematic constraint is obtained by writing the invariant mass of the
emitted neutrino as
\begin{equation}
   p^2_{\nu} = ( p_B - p_{D^*} - p_\ell )^2  \label{eqn:a}
\end{equation}
where $p$ stands for the 4-vector of the particle in subscript.
Expanding this equation results in
\begin{equation}
   p^2_{\nu} =
(E_B - E_{D^*\ell})^2 - |{\bf p}_B|^{~2} - |{\bf p}_{D^*\ell}|^{~2}
                   + 2|{\bf p}_B||{\bf p}_{D^*\ell}|cos\Theta    \label{eqn:b}
\end{equation}
where
$(E_B,{\bf p}_B)$ is the $B$ meson 4-momentum,
$(E_{D^*\ell},{\bf p}_{D^*\ell})$
is the sum of
the $D^*$ and lepton 4-momenta, and $\Theta$ is the angle between the
3-momenta ${\bf p}_{D^*\ell}$ and ${\bf p}_B$. The first three terms on the
right hand side constitute
what is traditionally referred to as missing mass squared, symbolized $MM^2$.
The factor multiplying $\cos\Theta$ will be denoted by $C$ for
{\em cosine multiplier}:
\begin{eqnarray}
    MM^2 &\equiv&
(E_B-E_{D^*\ell})^2 - |{\bf p}_B|^{~2}-|{\bf p}_{D^*\ell}|^{~2}, \\
    C    &\equiv& 2|{\bf p}_B||{\bf p}_{D^*\ell}|.
\end{eqnarray}
$E_{D^*\ell}$ and ${\bf p}_{D^*\ell}$
are determined from the measured momenta of the lepton and
$D^*$ candidates.  Since we
know the magnitude of ${\bf p}_B$, but not its direction,
$\cos\Theta$ is the only unknown.

\begin{figure}
\centerline{\psboxto(6.0in;6.0in){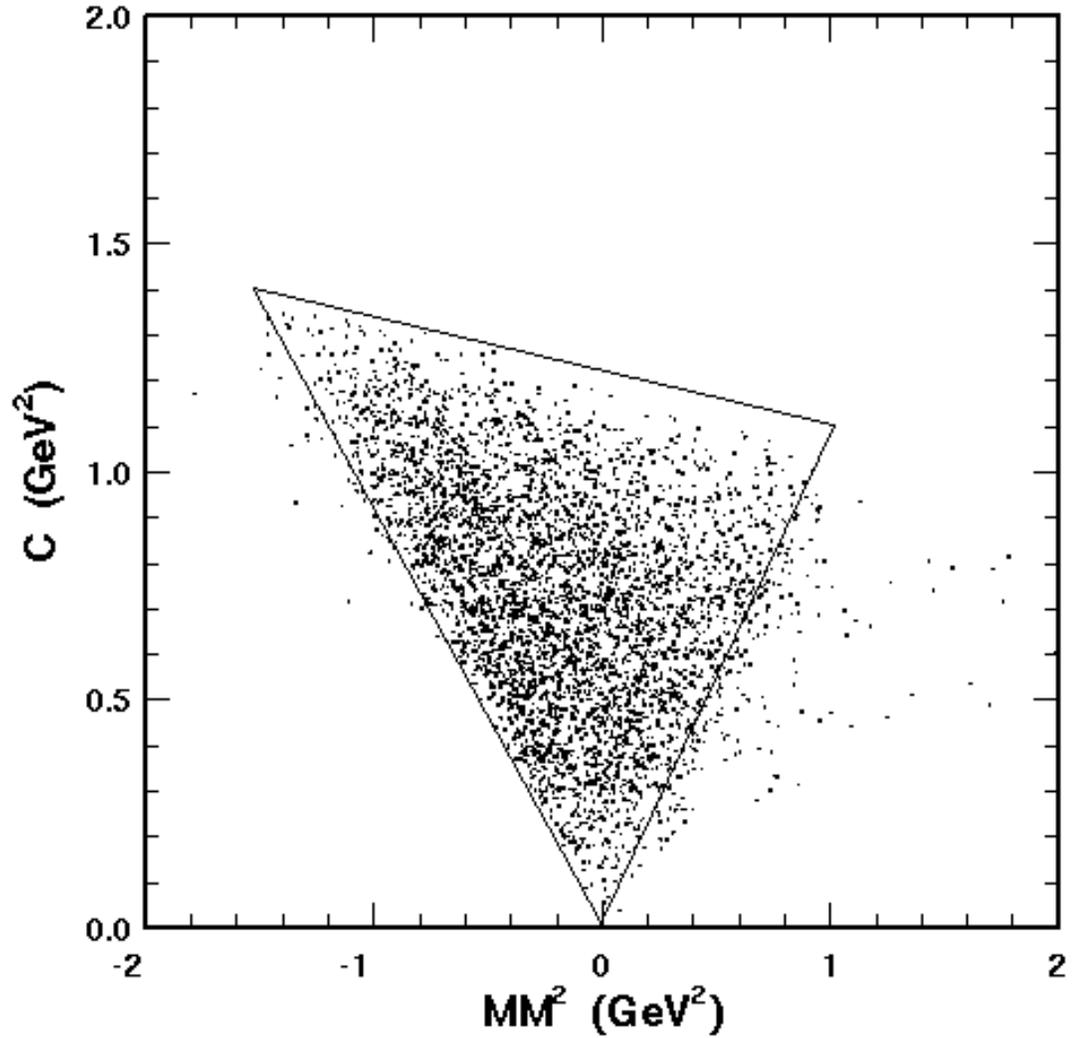}}
 \caption{Kinematic boundary for
$\bar{B} \rightarrow D^*~\ell~\bar{\nu}$ decays in the plane of
$C$ vs. $MM^2$, for lepton momentum in the range
$1.4 < |{\bf p}_\ell| < 2.4$~GeV (solid line).
This is the Signal Region for this analysis. The dots,
including those outside the kinematic boundary, are Monte Carlo signal
events.
Final state radiation and bremsstrahlung  occasionally force reconstructed
events outside of the kinematic boundary.}
  \label{fig:signalregion}
\end{figure}

For each $D^*$ and lepton combination in the same event
the pair of variables $C$ and $MM^2$ is calculated.
For correctly reconstructed $\bar{B} \rightarrow D^*~\ell~\bar{\nu}$ decays
(with perfect detector resolution),
the values of $C$ and $MM^2$ must lie within the
kinematic boundary determined by Eq.\ (\ref{eqn:b}) with
${\it p}^2_{\nu} = 0$. This boundary is
shown in Fig.\ \ref{fig:signalregion} for the lepton
momentum range  $1.4 \leq |{\bf p}_\ell| \leq 2.4$ GeV.
This lepton momentum range
and kinematic boundary define the Signal Region~\cite{signalregion}.
To arrive at the number of
true $\bar{B} \rightarrow D^*~\ell~\bar{\nu}$ decays in the data sample,
we count the number of candidates
observed in the Signal Region, subtract the expected number of
background candidates which happen to fall inside the Signal Region,
and divide by the Monte Carlo efficiency for signal events. Because
the Signal Region spans a significant area of phase space, we cannot assume
that backgrounds vary slowly in and near it, and a reliable
estimate of the background inside the Signal Region cannot be obtained
by interpolating from the number of candidates observed outside. Instead, we
categorize all sources of background and estimate the total contribution
from each source in the Signal Region by using data.  The method is
insensitive to the detailed $C$ vs. $MM^2$ distribution of the signals and
of all backgrounds except $\bar{B} \rightarrow D^*~X~\ell~\bar{\nu}$.

$\bar{B} \rightarrow D^*~X~\ell~\bar{\nu}$
decay is a significant source of background in this analysis,
principally because this is the background physical process most
similar to $\bar{B} \rightarrow D^*~\ell~\bar{\nu}$.
It includes resonant $\bar{B} \rightarrow D^{**}~\ell~\bar{\nu}$
decays followed by $D^{**} \to D^*X$, as well as non resonant decays.
This background is estimated from data events in a different region
of $C$ vs. $MM^2$ with $0.8< |{\bf p}_\ell |<1.4$~GeV. This region in $
|{\bf p}_\ell |$, $C$
and $MM^2$ is called the Correlated Background Region. It is described in
more detail in Section IVB, along with the method for subtracting
$\bar{B} \rightarrow D^*~X~\ell~\bar{\nu}$
background. The method is sensitive to the detailed $C$ vs. $MM^2$
distribution of
$\bar{B} \rightarrow D^*~X~\ell~\bar{\nu}$ background,
for which we use theoretical model predictions
(and to which we assign conservative errors).
Since this method must provide an estimate of the
number of entries in the Signal Region due to
$\bar{B} \rightarrow D^*~X~\ell~\bar{\nu}$ decay,
it also provides a model dependent measurement of
${\cal B}(\bar{B} \rightarrow D^*~X~\ell~\bar{\nu})$.


\section{Event Reconstruction}


The data used in this analysis were produced in symmetric
electron-positron collisions at the Cornell Electron-positron Storage
Ring (CESR) and recorded with the CLEO II detector.  The signal comes from
an integrated luminosity of 1.55~fb$^{-1}$ collected at the $\Upsilon(4S)$
center-of-mass energy \cite{lumin}.  An additional
0.69~fb$^{-1}$ of data collected below the $B\bar{B}$ production
threshold are used for continuum background determination.

The most crucial components of the CLEO II detector in this analysis are
the tracking system, the CsI electromagnetic calorimeter, and the muon
identification system.  A detailed description of the CLEO II detector
is given elsewhere \cite{detector}. The tracking system comprises a set of
 drift and straw tube chambers in a 1.5 Tesla magnetic field that measure
the momenta of stable charged particles over approximately $92\%$
of $4\pi$ with a transverse
momentum resolution  of $(\delta p_t/p_t)^2 = (0.0015p_t)^2 +
(0.005)^2$, where $p_t$ is measured in GeV.
Photons are detected in a CsI
 electromagnetic calorimeter with an angular acceptance of 95$\%$ of $4\pi$.
We restrict the fiducial volume for photons
to the barrel portion of the calorimeter, $|\cos \theta_\gamma| < 0.8$,
where $\theta_\gamma$ is the angle a photon makes with the beam line
(polar angle). The calorimeter
energy resolution is $\Delta E/E (\%) = 0.35/E^{3/4} + 1.9 - 0.1 E$,
where $E$ is in GeV \cite{badbarrel}, which corresponds to $4\%$ at 0.1~GeV.
Both electron and muon candidates
must lie within the polar angular region $|\cos\theta_\ell | < 0.71$.
In this analysis, electrons with momenta above 0.8 GeV
are identified by their electromagnetic
interactions in the calorimeter,
their energy loss in the drift chamber gas
and their time of flight in the detector.
The electron identification efficiency within the fiducial volume
is over $94\%$, while a hadron in the momentum range
0.8 to 2.4 GeV has on average
a $(0.3\pm 0.1)\%$ probability of being misidentified as an electron.
Muons are
identified by their ability to penetrate
at least 5 nuclear absorption lengths in iron, which puts a lower limit
of 1.4 GeV on the muon momentum acceptance.
Muons within the acceptance are identified
with 93\% efficiency, while  hadrons have on
average a $(1.4\pm0.2)\%$ probability of being misidentified as a muon.

For this analysis we select hadronic events \cite{hadronic} that have
at least one track identified as a lepton with momentum
$0.8 \leq |{\bf p}_\ell| \leq 2.4$ GeV.
The ratio of the second to the
zeroth Fox-Wolfram moments \cite{foxwol}
of the event is required to be less than $0.4$ to
suppress background from continuum events.
For each lepton in these events we search for $D^0$ candidates
in the decay mode $D^0 \rightarrow K^-\pi^+$
using charge correlation with the lepton to make unambiguous mass
assignments (the lepton and kaon charges must be the same).

We combine $D^0$ candidates with pion candidates to fully reconstruct
$D^*$ mesons
in the modes $D^{*+}\rightarrow D^0\pi^+$ and
$D^{*0}\rightarrow D^0\pi^0$. We  call these pion candidates
slow pions  because their momentum is restricted to be
less than 225~MeV in the laboratory frame.
Charged slow pions are accepted if they lie in the polar angle
region  $|\cos\theta_{\pi^+}|<0.71$ and have momentum above
65~MeV. Candidate $\pi^0$'s are constructed from
pairs of showers in the electromagnetic calorimeter which do not match
the projection of any drift chamber track and have  an invariant mass within
$3 \sigma$ of the measured $\pi^0$ mass ($\sigma =$ 5 to 8~MeV,
depending on shower energies and polar angles).
Showers used in $\pi^0$ candidates must be in the polar angle region
$|\cos\theta_{\gamma}|<0.8$ and have an energy above $30$~MeV. The
$\pi^0$ momentum vector is reconstructed by constraining the shower position
 and energy measurements to produce the known $\pi^0$ mass.
The momentum of $D^*$ candidates must satisfy $|{\bf p}_{D^{*}}|/
\sqrt{E_B^2 - m_{D^{*}}^2}<0.5$ to be consistent with $B$ decay.

The raw yield of events with a $D^*$ meson and a lepton is obtained by
fitting the $D^0$ mass peak after cutting on  the
$D^* - D^0$ candidate mass difference,
$\delta_m \equiv M_{K\pi\pi_s} - M_{K\pi}$, and
subtracting the scaled result of a $D^0$ mass fit to a
$\delta_m$ sideband ($M_{K\pi}$ and $M_{K\pi\pi_s}$
denote the invariant masses
of the $D^0$ and $D^*$ candidates, respectively).
This sideband subtraction is described in detail in the next section.


\section{Backgrounds}


The number of $D^*$ and lepton pairs observed in a given region of the $C$
vs. $MM^2$ plane is  the sum of the signal and background sources in that
region. The background sources fall into five categories, and we
have evaluated the contribution from each of these  sources
using the data. In order of importance the five background categories are:
combinatoric, correlated, uncorrelated, continuum, and fake lepton background.

\subsection{Combinatoric Background}

The dominant background in this analysis is the combinatoric background
in $D^{*+}$ and $D^{*0}$ reconstruction. One class of background arises
from combinations of random $K^-\pi^+$ pairs with any slow
second pion, $\pi_s$.  This class of background
does not peak in $M_{K\pi}$, but a subclass
of these events in which the $\pi_s$ and either the $K^-$ or $\pi^+$ are
from a true $D^* \to D^0 \pi$ decay does peak in $\delta_m$.
Combinations of $K^-\pi^+$
pairs from a correctly reconstructed $D^0$ meson with an unrelated
pion candidate forms a second class
of combinatoric background which peaks in $M_{K\pi}$, but not in $\delta_m$.
The $M_{K\pi}$
distributions in the $\delta_m$ signal and sideband regions
are fit to a $D^0$ peak plus a Chebyshev polynomial, which removes the first
class of background in both regions.  To remove the second class and obtain
a final signal yield, the $D^0$ yield in the $\delta_m$ sideband is
scaled and subtracted from the yield in the $\delta_m$ signal region. The
scale factor is determined from the background function in a fit to the
$\delta_m$ distribution~\cite{delmfits}.

\begin{figure}
\centerline{\psboxto(6.0in;6.0in){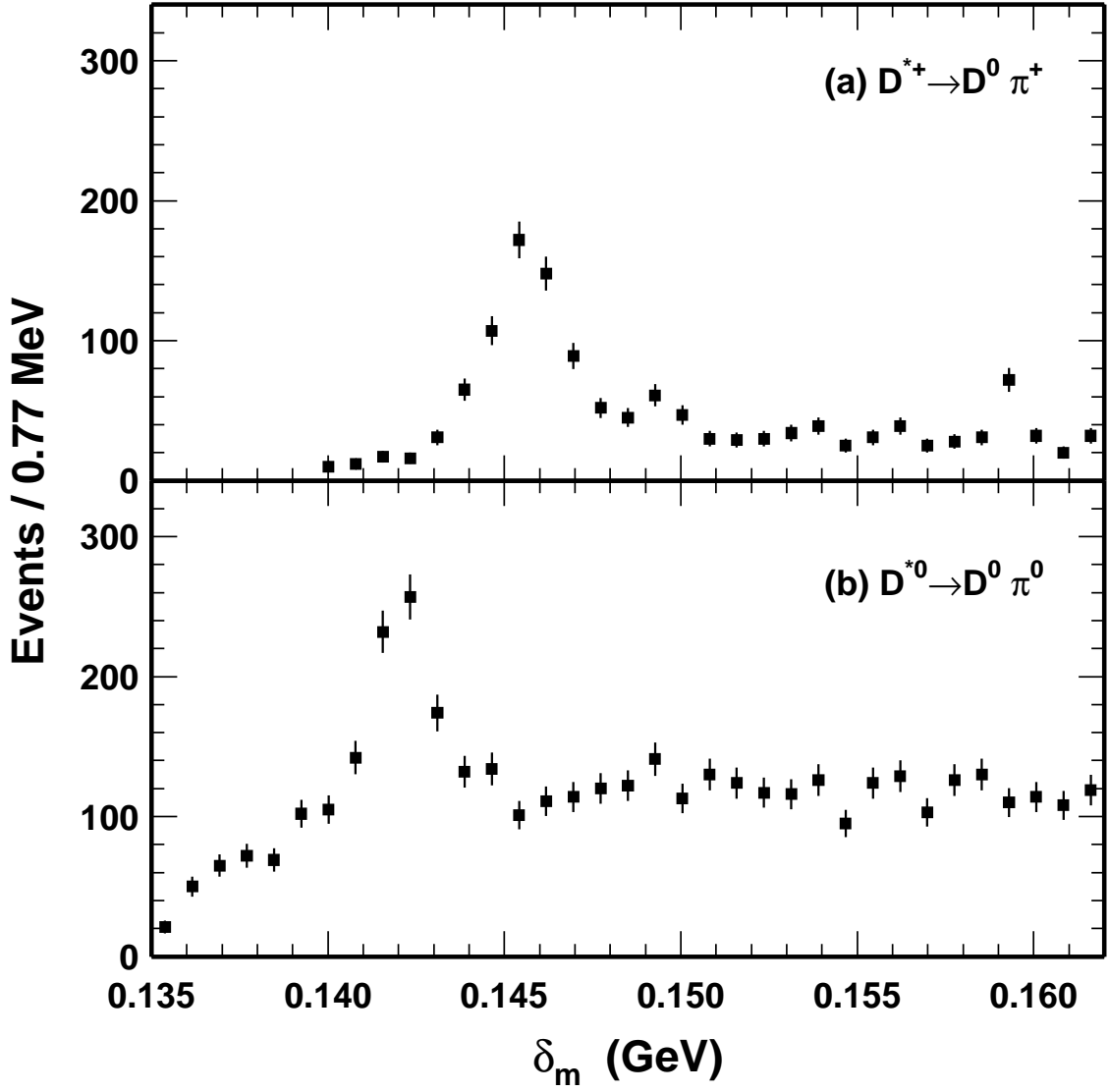}}
 \caption{Mass difference ($\delta_m$)
distributions of events falling in the Signal
Region of the decay modes (a)
$\bar{B}^0 \rightarrow D^{*+}~\ell^-~\bar{\nu}$ and
(b) $B^- \rightarrow D^{*0}~\ell^-~\bar{\nu}$. All candidates are required to
have $M_{K\pi}$ within 100~MeV of the measured $D^0$ mass.}
  \label{fig:deltam}
\end{figure}

The distributions of $\delta_m$ for $D^{*+}$ and $D^{*0}$ are shown in
Fig.\ \ref{fig:deltam} for events which lie in the $C$ vs. $MM^2$ Signal
Region and have $M_{K\pi}$ within 100~MeV of the measured $D^0$ mass.
The $\delta_m$ signal region for $D^{*+}$ is 8~MeV wide and
is centered on the measured $\delta_m$ mean of $145.44$~MeV
\cite{pdg}, while
the sideband region is $150<\delta_m<165$~MeV.
The $\delta_m$ signal region for $D^{*0}$ is 6~MeV wide and
is centered on the measured $\delta_m$ mean of $142.12$~MeV
\cite{isomas}, while
the sideband region is $147<\delta_m<162$~MeV. We have used
a wider signal
region for $D^{*+}$ because the $\delta_m$ signal in this mode
has non Gaussian tails from systematic effects in the
reconstruction of very low
momentum charged tracks. These effects are understood and reproduced in the
Monte Carlo simulation.

\begin{figure}
\centerline{\psboxto(6.0in;6.0in){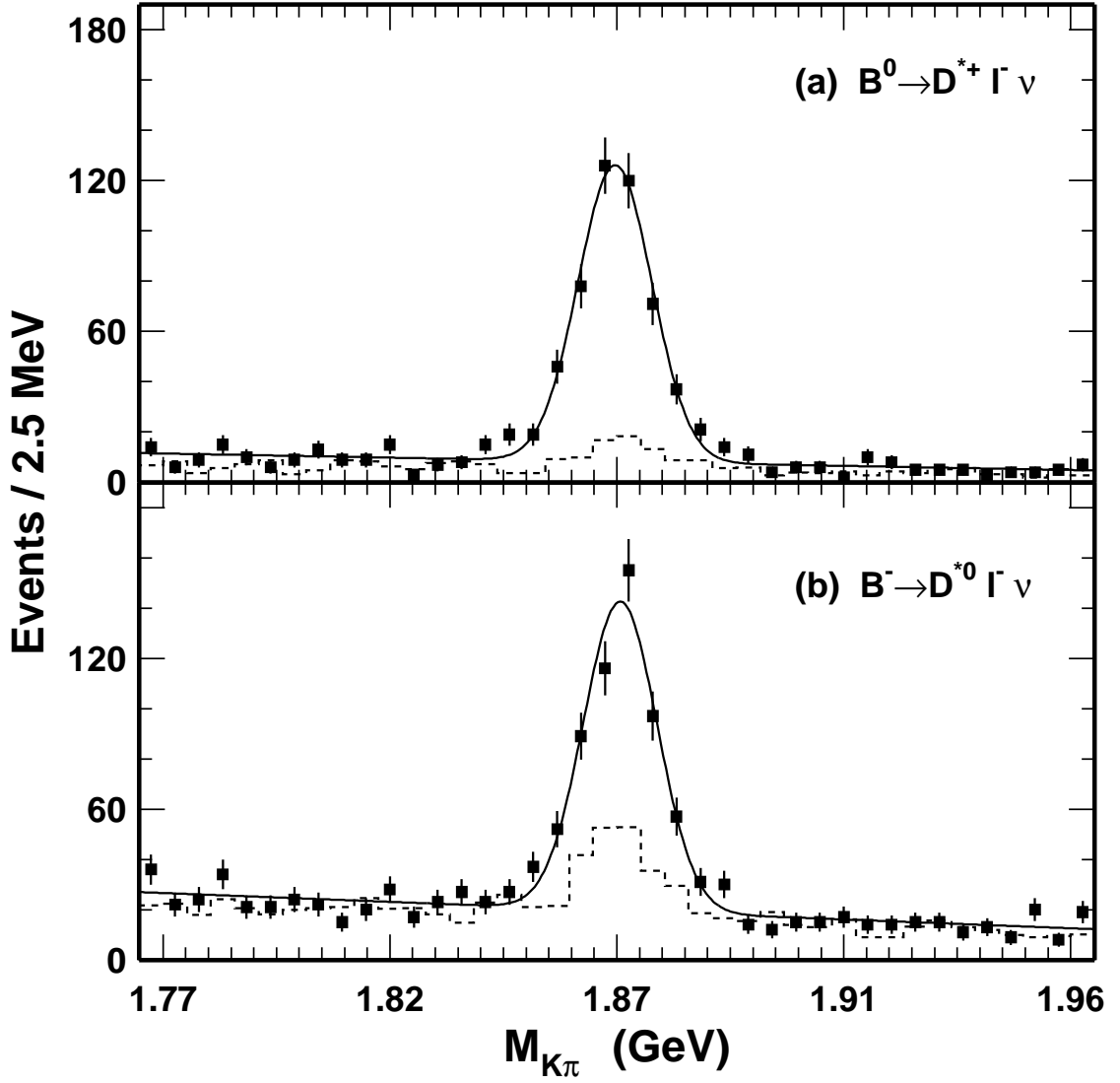}}
 \caption{Invariant mass of $K^-\pi^+$ combinations passing a $\delta_m$ cut
in the Signal Region of the decay modes (a)
$\bar{B}^0 \rightarrow D^{*+}~\ell^-~\bar{\nu}$ and
(b) $B^- \rightarrow D^{*0}~\ell^-~\bar{\nu}$.
The dashed histograms show the scaled $M_{K\pi}$
distributions of events in the Signal Region,
but in the $\delta_m$ sideband. The solid lines show the fits used to determine
the yields. }
  \label{fig:dsignals}
\end{figure}

The $M_{K\pi}$ distributions for events in the
$\delta_m$ signal and sideband regions are shown in Fig.\ \ref{fig:dsignals}.
These are the four distributions that were fit to obtain the yields in the
$C$ vs. $MM^2$ Signal Region. The yields are listed separately for
the $\delta_m$ signal and sideband regions in Tables \ref{table:numbers0}
and \ref{table:numbers-}.

\subsection{Correlated $D^*$--lepton Background}

\begin{figure}
\centerline{\psboxto(6.0in;6.0in){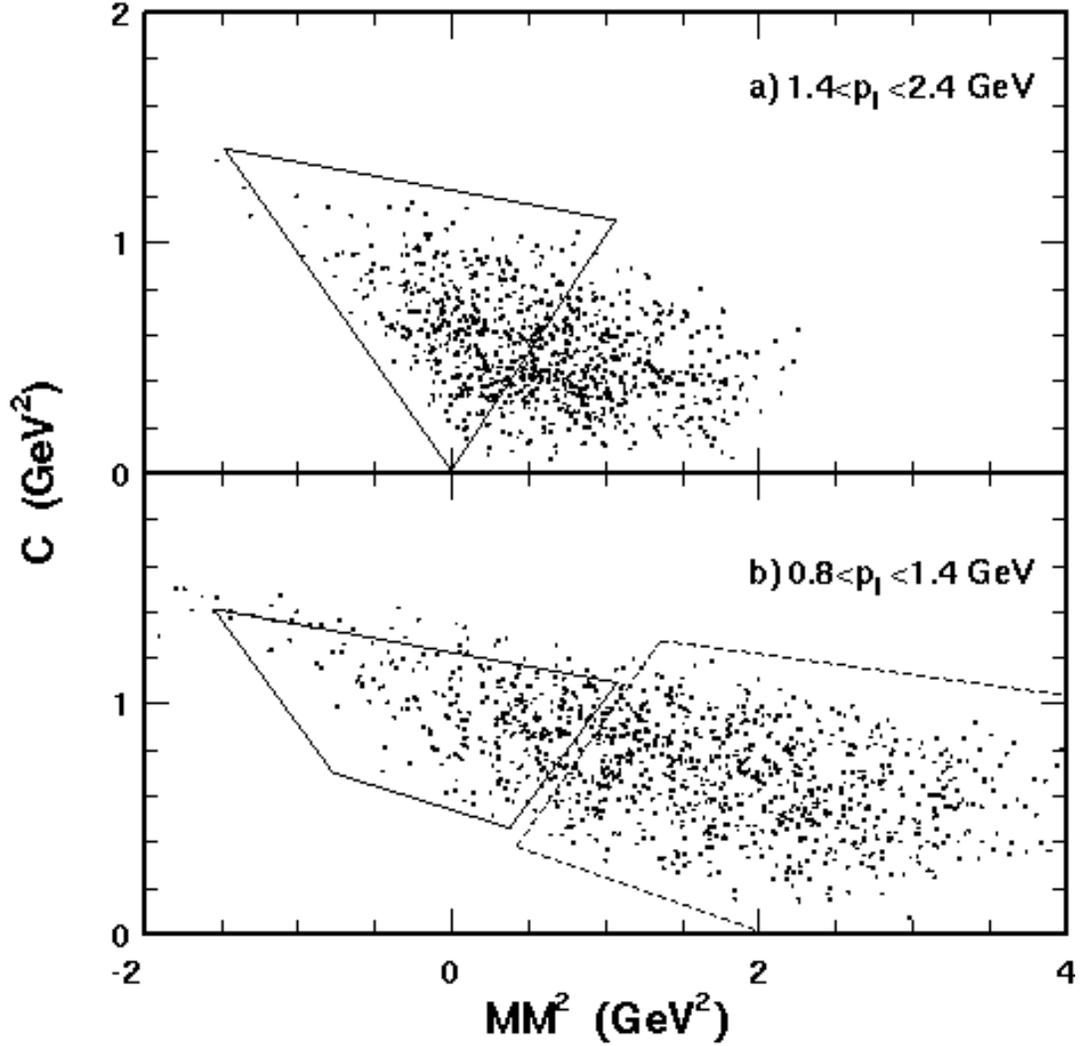}}
 \caption{Distribution of Monte Carlo
$\bar{B} \rightarrow D^{**}~\ell~\bar{\nu}$ events in the plane
of $C$ vs. $MM^2$, (a) for lepton momentum in the range
$1.4<|{\bf p}_\ell|<2.4$~GeV
and (b) $0.8<|{\bf p}_\ell|<1.4$~GeV.
The kinematic boundary for $\bar{B} \rightarrow D^*~\ell~\bar{\nu}$ decays
is also shown in each plot (solid line). The dashed line in the right plot
indicates the boundary of  the Correlated Background Region.}
  \label{fig:doubleregion}
\end{figure}

A $\bar{B}$ meson can decay to a final state with a
$D^*$ and an $e^-$ or $\mu^-$
through  channels other than
$\bar{B} \rightarrow D^*~\ell~\bar{\nu}$,
and these physical processes contribute
a background that we call correlated. The
principal source of correlated background is the decay
$\bar{B} \rightarrow D^*~X~\ell~\bar{\nu} $,
where e.g., $X = \pi$ or other unreconstructed particle(s) ($X \neq 0$).
To remove this background, we exploit
two general features of $\bar{B} \rightarrow D^*~X~\ell~\bar{\nu}$ decays.
First, because we do not reconstruct $X$
the $MM^2$ variable will tend to be shifted
towards positive values. Second, because the $D^*X$ invariant mass
is larger than the $D^*$ mass, the lepton spectrum will be softer than for
$\bar{B} \rightarrow D^*~\ell~\bar{\nu}$ decays.
Fig.\ \ref{fig:doubleregion} shows the $C$ vs. $MM^2$
distribution in the two lepton momentum ranges
$0.8< |{\bf p}_\ell|<1.4$~GeV and
$1.4<|{\bf p}_\ell|<2.4$~GeV for  Monte Carlo
$\bar{B} \rightarrow D^{**}~\ell~\bar{\nu}$ decays generated
according to the model of ISGW~\cite{isgw} (we take these
events to be representative, at the level of precision required here, of
generic $\bar{B} \rightarrow D^*~X~\ell~\bar{\nu}$ decays).
In the lower lepton momentum range there is a region of the $C$ vs. $MM^2$
plane where the $\bar{B} \rightarrow D^{**}~\ell~\bar{\nu}$
efficiency is high, but there is
almost no contribution from $\bar{B} \rightarrow D^*~\ell~\bar{\nu}$ decays.
Therefore, the low lepton momentum range is  used to measure the level
of $\bar{B} \rightarrow D^*~X~\ell~\bar{\nu}$,
which is then scaled using Monte Carlo to determine
the $\bar{B} \rightarrow D^*~X~\ell~\bar{\nu}$
contribution to the high lepton momentum Signal Region~\cite{cbr}.
We do not determine the level of $\bar{B} \rightarrow D^*~X~\ell~\bar{\nu}$
from the events just outside the high momentum Signal Region, because
the $\bar{B} \rightarrow D^{**}~\ell~\bar{\nu}$
efficiency is low in the high lepton momentum range and
electron bremsstrahlung in the detector causes
a small fraction of $\bar{B} \rightarrow D^*~\ell~\bar{\nu}$
decays to be reconstructed outside the
signal boundary (Fig.\ \ref{fig:signalregion}).

\begin{figure}
\centerline{\psboxto(6.0in;6.0in){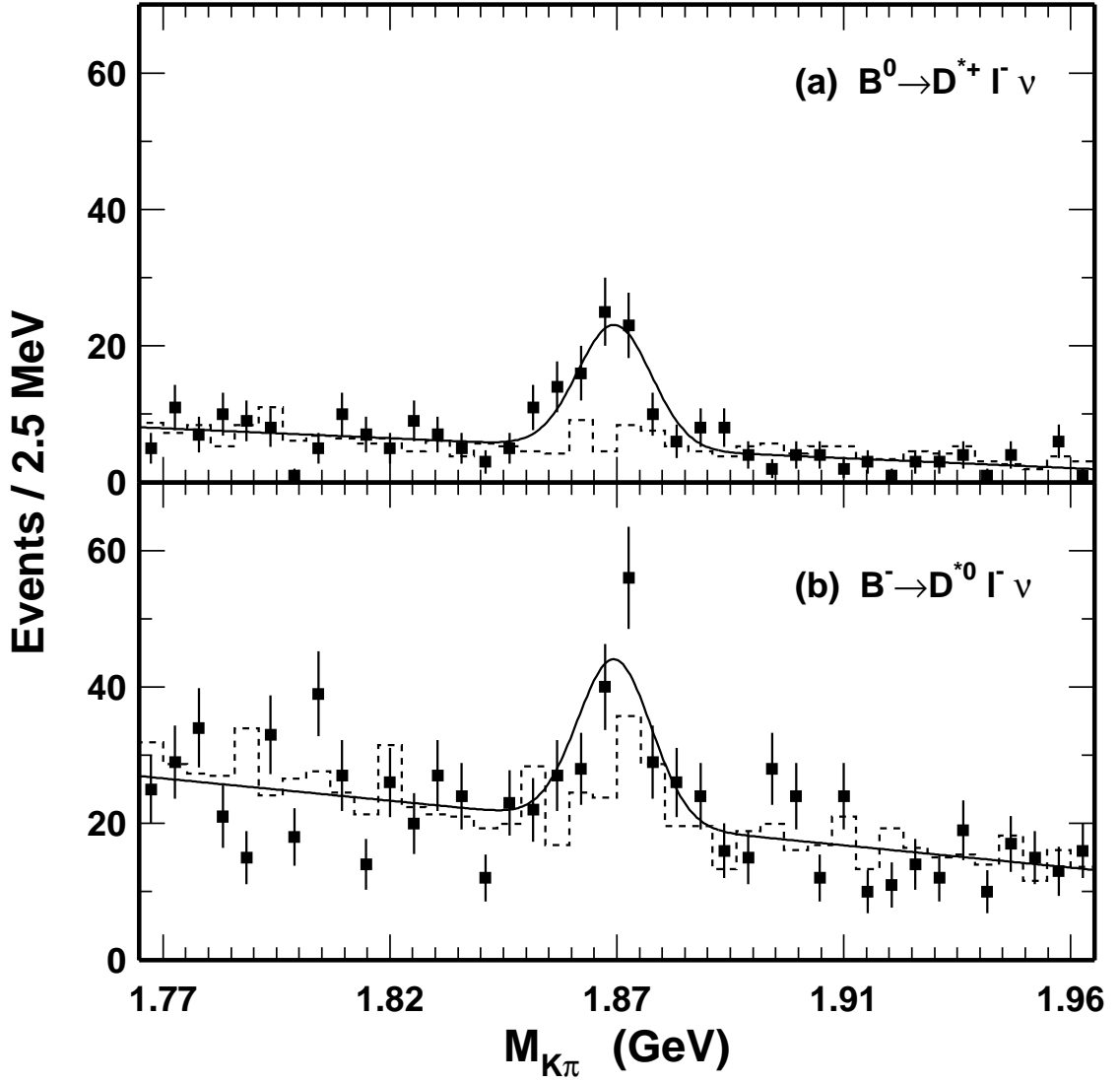}}
 \caption{Invariant mass of $K^-\pi^+$ combinations passing a $\delta_m$ cut
in the Correlated Background Region of the decay modes (a)
$\bar{B}^0 \rightarrow D^{*+}~\ell^-~\bar{\nu}$ and
(b) $B^- \rightarrow D^{*0}~\ell^-~\bar{\nu}$.
The dashed histograms show the scaled $M_{K\pi}$
distributions of events in the $\delta_m$ sideband.
The solid lines show the fits used to determine the yields. }
  \label{fig:doublesignals}
\end{figure}

The dashed lines in Fig.~\ref{fig:doubleregion} define a region in the $C$
vs. $MM^2$ plane for the lepton momentum range
$0.8<|{\bf p}_\ell|<1.4$~GeV called the Correlated Background Region,
because it has
high efficiency for the
$\bar{B} \rightarrow D^*~X~\ell~\bar{\nu}$
background processes, but very low efficiency for
our signal mode. The $M_{K\pi}$ distributions
in the Correlated Background Region are shown in Fig.\ \ref{fig:doublesignals}.
Let $N_S$ and $N_C$ denote the total yields in
the high lepton momentum Signal Region and the low lepton momentum Correlated
Background Region, respectively, after
subtraction of all other backgrounds as discussed in the following
sections.  They are related to $N_s(\bar{B} \rightarrow D^*~\ell~\bar{\nu})$,
the number of $\bar{B} \rightarrow D^*~\ell~\bar{\nu}$
events in the high momentum Signal Region,
and $N_c(\bar{B} \rightarrow D^*~X~\ell~\bar{\nu})$,
the number of $\bar{B} \rightarrow D^*~X~\ell~\bar{\nu}$ events in the low
momentum Correlated Background Region, via
\begin{equation}
   \left.
     \begin{array}{llrrrl}
        N_S & = & N_s(\bar{B} \rightarrow D^*~\ell~\bar{\nu})
& + & r_c N_c(\bar{B} \rightarrow D^*~X~\ell~\bar{\nu})  & ~ \\
        N_C & = & r_s N_s(\bar{B} \rightarrow D^*~\ell~\bar{\nu})
& + & N_c(\bar{B} \rightarrow D^*~X~\ell~\bar{\nu})     & ~
      \end{array}
   \right\} \label{eqn:simultaneous}
\end{equation}
where $r_s = 0.015\pm0.003$ is the ratio of the
efficiency for $\bar{B} \rightarrow D^*~\ell~\bar{\nu}$  events
in the Correlated Background Region to the
efficiency in the Signal Region, and
$r_c = 0.70 \pm0.35$ is the ratio of the
$\bar{B} \rightarrow D^*~X~\ell~\bar{\nu}$ efficiency in the signal
region to the efficiency in the Correlated Background
Region.   Even though the Correlated Background Region is outside the nominal
kinematic boundary for signal, $r_s$ is non zero because of electron
bremsstrahlung.  The values for $r_s$ and $r_c$ were determined from Monte
Carlo simulation, and the errors reflect uncertainties due to model dependence.
We estimated $r_c$ using resonant
$\bar{B} \rightarrow D^{**}~\ell~\bar{\nu}$ decays.
These were generated according to the ISGW model \cite{isgw} for
the $^1P_1$, $^3P_1$, and $^3P_2$ $D^{**}$ states in their
predicted relative abundances \cite{scora}. We expect that the
$B \rightarrow D^{**}(^3P_1) \ell\bar{\nu}$ decay
is a reasonable approximation to non-resonant
$B \rightarrow (D^{*}\pi) \ell\bar{\nu}$
decays because the $^3P_1$ is expected to be a
very wide resonance. We take into account our rough modelling of
$\bar{B} \rightarrow D^*~X~\ell~\bar{\nu}$
decays by assigning a conservative error of $50\%$ to $r_c$.

The observed numbers of events and the yields obtained by solving
Eq.\ (\ref{eqn:simultaneous}) with the quoted central values of $r_s$ and $r_c$
are summarized in Tables~\ref{table:numbers0} and~\ref{table:numbers-}. The
uncertainties in $r_s$ and
$r_c$ are included in the systematic errors for the final
$\bar{B} \rightarrow D^*~\ell~\bar{\nu}$ and
$\bar{B} \rightarrow D^*~X~\ell~\bar{\nu}$ yields.

Sources other than $\bar{B} \rightarrow D^*~X~\ell~\bar{\nu}$
contribute to the correlated background
at much smaller levels.
The decays $B \to D^*X$, where $X$ fragments to light hadrons that
decay semileptonically or are misidentified as leptons, are discussed
in the section on lepton fakes. The processes
$B \to D^* \tau^- \bar{\nu}~$ followed by $\tau^- \to l^-\bar{\nu}\nu~$, and
$B \to D^* D_{(s)}^{(*)-}~$ followed by $D_{(s)}^{(*)-} \to X l^-\bar{\nu}$
lead to final states that include a $D^*$ and a lepton with the
same ``right sign'' charge
correlation as signal. These sources have characteristics similar to
$\bar{B} \rightarrow D^*~X~\ell~\bar{\nu}$,
but with an even softer lepton spectrum. Monte Carlo
simulation predicts values for $r_c$ of
$0.01\pm0.01$ and $0.005\pm0.005$ for
$B \to D^* \tau^- \bar{\nu}~$
and $B \to D^* D_{(s)}^{(*)-}~$ respectively. They can therefore be
neglected in the Signal Region. We have estimated their contributions to
the background region using available measurements or estimates of the various
intermediate branching fractions \cite{pdg,btaunu} together with  efficiencies
determined in Monte Carlo simulations. The results are given in Tables
\ref{table:numbers0} and \ref{table:numbers-}.

\subsection{Uncorrelated $D^*$ lepton background}

``Uncorrelated" background encompasses events with  a $D^*$ from the
decay of the $\bar{B}$
and a lepton from the $B$ in a $B\bar{B}$ event. However,
in order to contribute to the background
in this analysis the $D^*$ and the lepton must have the same charge
correlation as  signal decays.
Therefore, the leptons in uncorrelated background are
secondary $B$ meson decay products from the decay chain $\bar{b} \to \bar{c}
\to \ell$ (known as cascades) or $B^0\rightarrow \bar{B}^0
\rightarrow \ell X$ (mixing).
It is also possible to have leptons from the decay
or misidentification of light hadrons ($K$ or $\pi$).

The cross section for producing uncorrelated background events
is measured using the data.
Because the $B$ and $\bar{B}$ mesons in an event decay
independently, the uncorrelated background cross section is a product
of the inclusive $D^*$ cross section from $\Upsilon(4S)$ decay,
$\sigma'(e^+e^- \to \Upsilon(4S) \to D^*)$, and the $B \to \ell^-$
inclusive branching fraction, ${\cal B}'(B \to \ell^-)$. The primes indicate
that the quantities of interest are the {\em detected} (``raw") cross
section and
branching fraction, which include detector acceptance
and reconstruction efficiencies as well as the underlying cross section
and branching fraction.
We measure these two quantities with the same event
selection and cuts used to obtain
$\bar{B} \rightarrow D^*~\ell~\bar{\nu}$ yields.

We find that the cross sections for producing and detecting a $D^*$
from $\Upsilon(4S)$ decay
(integrated over momentum) are $3.1\pm0.15$~pb and $3.2\pm0.33$~pb for
$D^{*+}$ and $D^{*0}$ respectively.

Measuring ${\cal B}'(B \to \ell^-)$ is not as straightforward as counting
leptons because of the need to preserve the correct $D^*$ and lepton charge
correlation, {\it i.e.}, we need specifically ${\cal B}'(B \to \ell^-)$, rather
than ${\cal B}'(\Upsilon(4S) \to \ell^{\pm})$.
We measure ${\cal B}'(B \rightarrow \ell^-)$
using like charge dilepton events,
in which one of the leptons tags one daughter
of the $\Upsilon(4S)$ as either a $B$ or a $\bar{B}$~\cite{beauty}.
We find that ${\cal B}'(B \to \ell^-)$ is
$(0.7\pm0.05)\%$ and $(1.2\pm0.05)\%$
in the momentum range $1.4< |{\bf p}_\ell | <2.4$ and $0.8<
|{\bf p}_\ell | <1.4$~GeV, respectively.
These detected branching fractions include many
sources, such as misidentified hadrons from the ``other'' $B$, photon
conversions, etc., but these all contribute to uncorrelated background and
therefore belong in the quantity we are trying to measure.

Multiplying $\sigma'(e^+e^- \to \Upsilon(4S) \to D^*)$ by
${\cal B}'(B \to \ell^-)$ yields the total cross section for uncorrelated
background. However, we wish to estimate
the number of uncorrelated background events in a particular region
of the $C$ vs. $MM^2$ plane; therefore the distribution of
uncorrelated background in these variables must be known.
These variables depend only on the inclusive $D^*$ and lepton
momentum distributions (which we have already measured)
and the angle between the $D^*$ and the lepton, $\alpha$.
The distribution of
$\cos\alpha$ is flat because the two $B$ mesons are produced
nearly at rest.  We use these three known distributions to simulate the
$C$ vs. $MM^2$ distribution of uncorrelated background.
{}From this simulation and the measured uncorrelated background cross section
we estimate the numbers of uncorrelated background events in the
Signal Region and Correlated Background Region (given in Tables
\ref{table:numbers0} and \ref{table:numbers-}).

\subsection{Continuum and Lepton Fakes}

The level of continuum background is estimated using the 0.69 fb$^{-1}$ of
data recorded at energies slightly below the $\Upsilon(4S)$ resonance.
These data are analyzed in the same manner as resonance data in order
to estimate the continuum backgrounds listed in Tables \ref{table:numbers0}
and \ref{table:numbers-}.

There is a small background of $D^* \ell$ candidates where the lepton is
actually a hadron that has been misidentified as a lepton.
We call this the fake lepton background.
Fake $D^* \ell$ pairs are predominantly uncorrelated, simply because
there is more energy available to produce light hadrons in the decay of the
$B$ meson that did not produce the $D^*$. These uncorrelated fake leptons
are already included in the cross section for uncorrelated background.

The correlated lepton fake contribution to the raw number of
$D^* \ell$ pairs is estimated by performing a
similar analysis of the data where
$D^*$ candidates are paired with light hadron candidates instead of leptons,
and scaling the result by the known electron and muon fake rates. For this
study a light hadron is defined as any detected
charged particle that fails very loose electron and muon identification cuts.
We use Monte Carlo simulation to determine the efficiency for correlated
$D^*$ and hadron pairs to fall within the signal and background regions of
the $C$ vs. $MM^2$ plane. The correlated lepton fake contributions
turn out to be so small that we can easily afford a large uncertainty due to
the use of Monte Carlo.
The results are given in Tables \ref{table:numbers0}
and \ref{table:numbers-}.


\section{Detection Efficiencies}
\label{sec:eff}
The efficiency for $\bar{B} \rightarrow D^*~\ell~\bar{\nu}$
events to pass our event selection criteria
is estimated from a Monte Carlo simulation.  The generator produces events
obeying all kinematic constraints due to angular momentum conservation and the
$V-A$ nature of the pseudoscalar to vector decay\cite{anders}.
These events are
passed through a full detector simulation \cite{geant} and then analyzed
as described above.

An uncertainty in the efficiency for
$\bar{B} \rightarrow D^*~\ell~\bar{\nu}$ exists because the dynamics of
the decay depend on unknown form factors~\cite{cleo:dstlnuff}.
Several phenomenological models of these form factors are available.
For the results presented here, we have used the
predictions of Neubert~\cite{neubert,neubertmodel}
to determine the central values
of our efficiencies, and have
compared these results to those obtained with the ISGW\cite{isgw},
BSW\cite{bsw}, and KS\cite{ks} models to estimate the
model uncertainty (See Tables ~\ref{table:modused} and
{}~\ref{table:vcbmod}).  After comparing these models as well as varying
the form factors within the Neubert model,
it is estimated that the model dependence
contributes a $3\%$ systematic uncertainty in the detection efficiency.

The efficiencies
for $\bar{B} \rightarrow D^*~\ell~\bar{\nu}$ events to be counted as signal are
\begin{eqnarray}
 \epsilon_s(\bar{B}^0 \rightarrow D^{*+}~\ell^-~\bar{\nu}) &=&
[9.54 \pm 0.23 \pm 0.71]\% \\
 \epsilon_s(B^- \rightarrow D^{*0}~\ell^-~\bar{\nu}) &=&
[7.18 \pm0.52 \pm0.53]\%
\end{eqnarray}
where the first (systematic) error is correlated between the two efficiencies
and the second is
uncorrelated. These
efficiencies do not include the $D^*$ and $D^0$ branching fractions.  The
errors include: Monte Carlo statistics, model dependence,
an uncertainty reflecting changes in efficiency corrected yields due to
variation of cuts,
the uncertainty in the normalization of the $\delta_m$ sidebands,
the uncertainty in the lepton reconstruction and identification, a 2\% per
track uncertainty in the reconstruction of $K$ and $\pi$ tracks
with momenta above 250 MeV, and the uncertainty in modelling the
efficiency of the slow pions. The methods used to estimate the latter
uncertainty are described below. The various
contributions to the systematic uncertainty in the efficiency are listed in
Table~\ref{table:efficiency}.

In the laboratory frame,
the slow pion from the
$\bar{B} \rightarrow D^*~\ell~\bar{\nu}$, $D^* \rightarrow D\pi$ decay chain
has a momentum of less than 225 MeV
with the peak of the distribution at 100 MeV.  The
efficiency to reconstruct $\pi^0$'s is fairly flat in this momentum range
but difficult to model because of the large background of low energy
showers in the calorimeter.  The efficiency for
reconstructing $\pi^+$'s decreases sharply below 100 MeV,
falling to 0 at 50 MeV.  It is crucial to
this analysis that these efficiencies be correctly reproduced by the Monte
Carlo and we have performed several detailed studies to evaluate the
reliability of our simulation~\cite{mgsthesis}.

\subsection{$\pi^+$ Efficiency}

The shape of the efficiency curve for charged pions of momentum less than
225 MeV can be measured using inclusive $D^{*+}$ decays in the data.
In the strong decay
$D^{*+} \rightarrow D^0 \pi^+$, $dN/d\cos\varphi_{D^*\pi}$ must be symmetric
about $\cos\varphi_{D^*\pi} = 0$, where $\varphi_{D^*\pi}$ is the angle
between the slow $\pi^+$ momentum in the $D^{*+}$ rest frame
and the $D^{*+}$ direction in the lab frame.  However, the observed
$dN/d\cos\varphi_{D^*\pi}$  distribution in the data may be asymmetric
because the slow $\pi^+$ efficiency varies with pion momentum,
which is highly correlated with $\cos\varphi_{D^*\pi}$.

\begin{figure}
\centerline{\psboxto(6.0in;6.0in){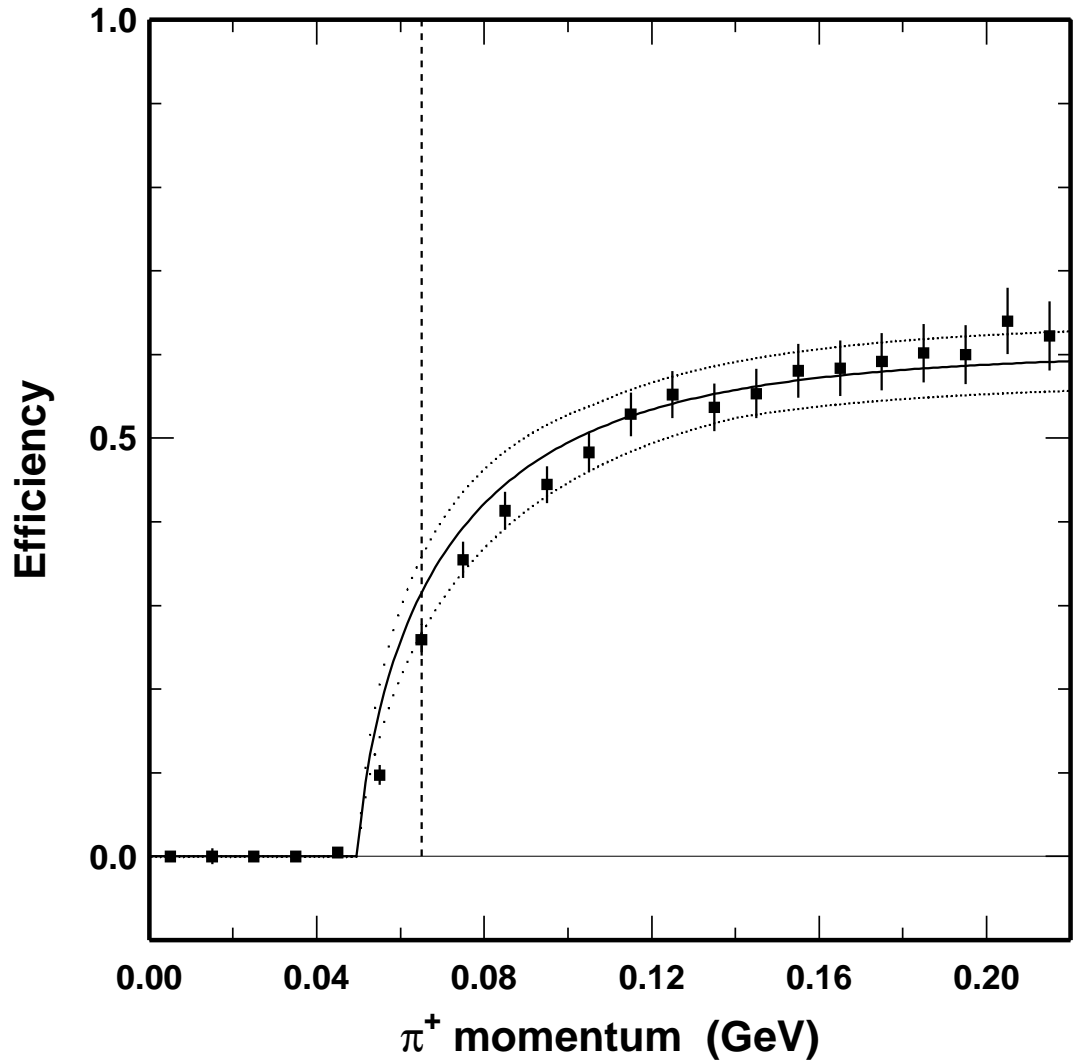}}
 \caption{Monte Carlo reconstruction efficiency versus momentum
(filled diamonds)
for charged slow pion candidates.
Overlaid (solid curve) is the efficiency as determined from the inclusive
$D^{*+}$ decay angle distribution in data. The $1\sigma$ variations in the
parameters of this curve are shown by the dotted lines. The dashed line
indicates the minimum momentum cut used in this analysis.
The efficiency shown here includes the geometric acceptance of 71\%.}
  \label{fig:pipeff}
\end{figure}

The charged pion efficiency between 0 and 225 MeV was measured in data by
simultaneously fitting the observed $dN/d\cos\varphi_{D^*\pi}$ distributions
in 8 bins of $D^{*+}$
momentum between 0 and 5 GeV. It was found that
the Monte Carlo reproduces the data
efficiency shape over the pion  momentum range 0 to 225 MeV
(Fig. \ref{fig:pipeff}). This study of the $\cos\varphi_{D^*\pi}$
distribution tests the simulation of the shape of the efficiency curve,
but not its
absolute normalization. We test the normalization by comparing
the yield of fully reconstructed $D^0 \rightarrow
K^-\pi^+\pi^0$ to the yield of partially reconstructed  $D^0 \rightarrow
K^-\pi^0(\pi^+)$, where the $\pi^+$ is not used. The momentum range
for these pions is $0.2$ to $1.0$~GeV.  The Monte Carlo
efficiency agrees well with the absolute $\pi^+$
efficiency measured in the data, within the statistical precision
of $2\%$. Together, these studies check at the $5\%$ level the simulation
of the slow $\pi^+$ efficiency.

\subsection{$\pi^0/\pi^+$ Efficiency}

The accuracy of the simulation of the slow $\pi^0$ efficiency
is difficult to test.  The
method outlined above cannot be used for the decay $D^{*0}\rightarrow D^0\pi^0$
because of the large combinatoric background in the $\pi^0$ reconstruction.
Instead, we compare the ratio of efficiencies,
$\epsilon(\pi^0)/\epsilon(\pi^+)$, in the pion
momentum range 0 to 225 MeV between data and Monte Carlo. We then combine
the results
of this study with the uncertainty in the  charged pion absolute efficiency
described above in order to
obtain the uncertainty in the $\pi^0$ absolute efficiency.

We determine the ratio of efficiencies below 225~MeV,
$c_1 = \epsilon(\pi^0)/\epsilon(\pi^+)$,
using a combination of two
methods.  The first part uses the decay $\eta \rightarrow \pi^+ \pi^- \pi^0$
as a source of both neutral and charged soft pions.  Candidate $\eta$'s are
selected with either a $\pi^0$ in the low (0 to 225 MeV) momentum
range and both charged pions above 250 MeV or a $\pi^{\pm}$
in the low momentum range
and the other charged and neutral pions with momentum above 250 MeV.
The ratio of the number of events in these two bins depends
on the efficiency to find the pions times a kinematic factor that depends
on the $\eta$ Dalitz plot and the $\eta$ momentum spectrum.  The kinematic
factor can be interpreted as the average probability for the three pions from
the $\eta$ decay to populate the desired momentum ranges.  The $\eta$ Dalitz
plot has been precisely studied in previous experiments \cite{eta}, and the
$\eta$ momentum spectrum in the Monte Carlo was tuned to the
data using the $\eta \rightarrow \gamma\gamma$ decay mode.
The ratio of yields is proportional to the ratio $c_1/c_2$ where $c_1$ is
defined above and $c_2$ is the ratio of neutral to charged
pion efficiencies above 250 MeV.
The ratio $c_1/c_2$ is determined in data and
compared with the same ratio calculated from Monte Carlo.

A second study using $K^0_S$'s then determines the product $c_1c_2$.
The $K^0_S$ study takes advantage of the fact that the ratio of
branching fractions ${\cal B}(K^0_S \rightarrow \pi^0\pi^0)/
{\cal B}(K^0_S \rightarrow \pi^+\pi^-)$ is well known.  Therefore,
the ratio of neutral to charged final state yields is the product
of a known ratio of branching fractions, an efficiency ratio and an acceptance
factor which is determined from Monte Carlo.  For kaons of
momentum 250 to 500~MeV, one daughter pion is in the range 60 to 225~MeV
and the other daughter is above 225~MeV so that the ratio of yields
$N(K^0_S \to \pi^0\pi^0)/N(K^0_S \to \pi^+\pi^-)$ in the $K^0_S$ momentum
range 250 to 500~MeV is proportional to the product $c_1c_2$.

\begin{figure}
\centerline{\psboxto(6.0in;6.0in){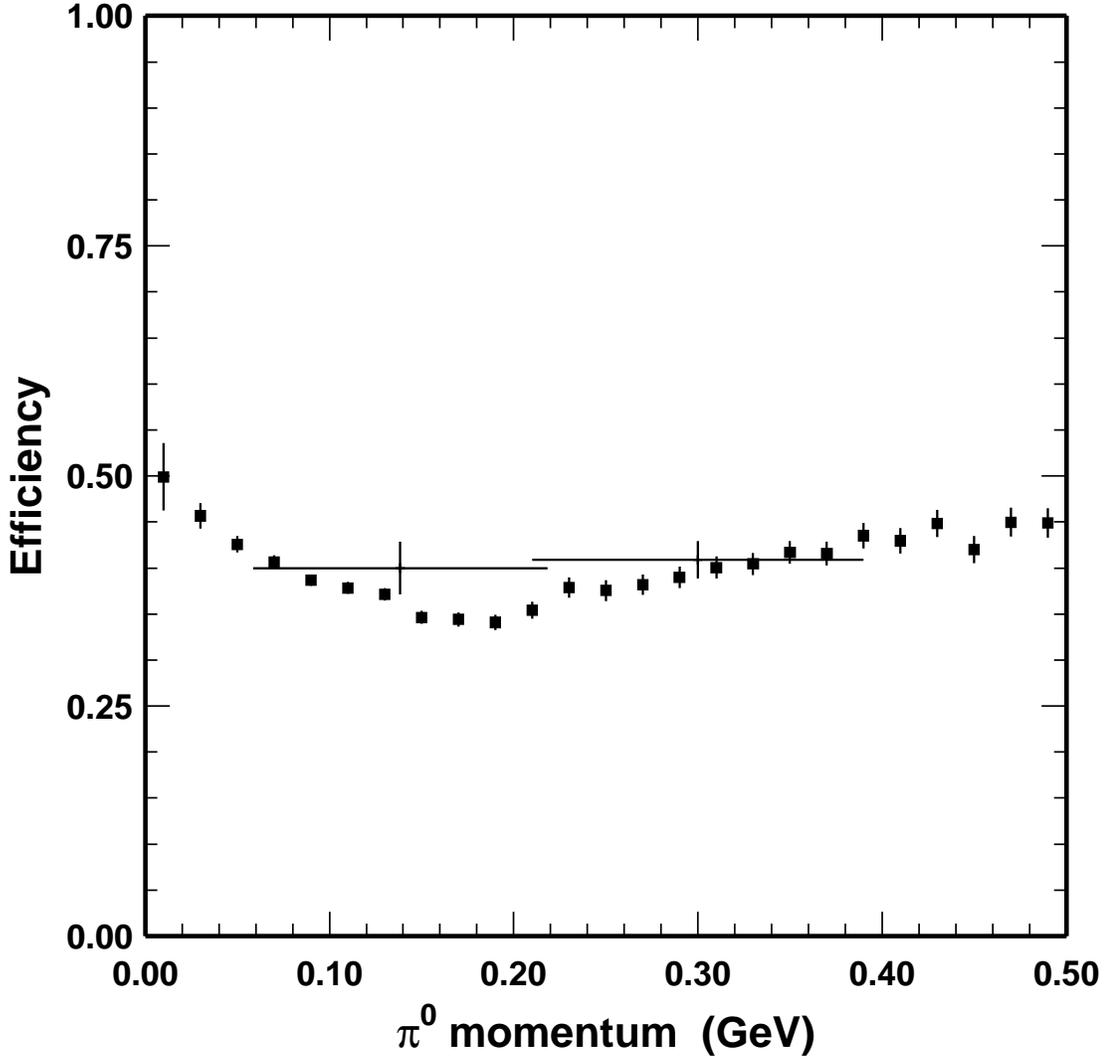}}
 \caption{Monte Carlo reconstruction efficiency versus momentum
(filled diamonds)
for neutral slow pion candidates.
Overlaid (crosses) are the results of the $\eta$ and $K^0_S$ studies which
compare the ratio of neutral to charged pion efficiency in data and Monte
Carlo. These two points were calculated assuming that the Monte Carlo
exactly simulates the charged pion efficiency.
The efficiency shown here includes the geometric acceptance (not constant with
$\pi^0$ momentum), which accounts for much of the inefficiency.}
  \label{fig:pizeff}
\end{figure}

When the results of the $\eta$  and $K^0_S$ studies are combined we find that
$c_1$ (the ratio of neutral to
charged pion efficiency for momentum 60 to 225 MeV), as
determined from the data, agrees with the predictions of the Monte Carlo to
within the statistical precision of $7\%$.
We therefore assign a systematic
error on the efficiency ratio of neutral to charged slow pions of 7\%.
Combining this
with our study of the absolute slow charged pion efficiency, we assign an
error of $8.6\%$ to the absolute determination of the neutral pion efficiency
in the momentum range below 225 MeV.  The Monte Carlo prediction of
the efficiency to reconstruct neutral pions
is shown in Fig. \ref{fig:pizeff}.

\section{Branching Fraction Results}

In this analysis we measure two independent quantities: $N_0$ and $N_-$ of
Eqs.\ (\ref{eqn:nzero}) and (\ref{eqn:nminus}).
These are in turn given by
\begin{eqnarray}
N_0 = { N_s(\bar{B}^0 \rightarrow D^{*+}~\ell^-~\bar{\nu})
\over \epsilon_s(\bar{B}^0 \rightarrow D^{*+}~\ell^-~\bar{\nu}) } \\
N_- = { N_s(B^- \rightarrow D^{*0}~\ell^-~\bar{\nu})
\over \epsilon_s(B^- \rightarrow D^{*0}~\ell^-~\bar{\nu}) }
\end{eqnarray}
where $N_s(\bar{B}^0 \rightarrow D^{*+}~\ell^-~\bar{\nu})$ and
$N_s(B^- \rightarrow D^{*0}~\ell^-~\bar{\nu})$
are the background subtracted yields
for the given signal modes
(Tables~\ref{table:numbers0} and~\ref{table:numbers-}),
and $\epsilon_s(\bar{B}^0 \rightarrow D^{*+}~\ell^-~\bar{\nu})$ and
$\epsilon_s(B^- \rightarrow D^{*0}~\ell^-~\bar{\nu})$ are the efficiencies for
the given signal modes (Section~\ref{sec:eff}).
Dividing $N_0$ and $N_-$ by four times
the number of $\Upsilon(4S)$ decays in the data sample
yields the product branching fractions given in Table~\ref{table:results}.
The number of $\Upsilon(4S)$ events was determined by studying the hadronic
cross section at energies both
on the $\Upsilon(4S)$ resonance and slightly below
the resonance in the continuum.
For our data sample, $N_{\Upsilon(4S)} = (1.65 \pm 0.01)
\times 10^6$ events.

The top two product branching fractions of Table~\ref{table:results}
depend on the
unknowns to be determined,
${\cal B}(\bar{B}^0 \rightarrow D^{*+}~\ell^-~\bar{\nu})$ and
${\cal B}(B^- \rightarrow D^{*0}~\ell^-~\bar{\nu})$, as well as the two
unmeasured quantities $f_{+-}$ and $f_{00}$. In all, there
are four unknowns and only two measurements.
The assumption  that
the $\Upsilon(4S)$ decays only to $B\bar{B}$
($f_{+-}+f_{00}=1$) reduces the
number of unknowns to three. This assumption is supported by the observation
that the $\Upsilon(4S)$ width is three orders of magnitude larger
than lower $\Upsilon$ states which are not massive enough to decay
to a pair of $B$ mesons \cite{besson}. There are two ways to further reduce
the number of unknowns.
The first method is to make the traditional assumption that
$f_{+-}=f_{00}$, which is uncertain at the $5-10\%$ level~\cite{lepage}.
Using this assumption we present the two independent
measurements:
${\cal B}(\bar{B}^0 \rightarrow D^{*+}~\ell^-~\bar{\nu})$ and
${\cal B}(B^- \rightarrow D^{*0}~\ell^-~\bar{\nu})$.
Alternatively, we can assume that
${\cal B}(B^- \rightarrow D^{*0}~\ell^-~\bar{\nu}) /
{\cal B}(\bar{B}^0 \rightarrow D^{*+}~\ell^-~\bar{\nu}) =
\tau_{B^-} / \tau_{\bar{B}^0}$.
This is equivalent to assuming that the partial widths for the exclusive
semi-leptonic decays of the charged and neutral $B$'s are equal:
$\Gamma(B^- \rightarrow D^{*0}~\ell^-~\bar{\nu}) =
\Gamma(\bar{B}^0 \rightarrow D^{*+}~\ell^-~\bar{\nu})$.
With this assumption and
a measurement of  $\tau_{B^-} / \tau_{\bar{B}^0}$ from other
experiments~\cite{lifetimes},
our measurement of $N_0$ and $N_-$ can be used to determine
$\Gamma(\bar{B} \rightarrow D^*~\ell~\bar{\nu})$ and $f_{+-} / f_{00}$.
We will first discuss the $\bar{B} \rightarrow D^*~\ell~\bar{\nu}$
branching fractions obtained using the two
different assumptions, and then the $\Upsilon(4S)$ branching fractions obtained
with the second assumption.

\subsection{$\bar{B} \rightarrow D^*~\ell~\bar{\nu}$ Branching Fractions}
\label{section:partial}
To extract the $B$ meson semileptonic branching fractions from
Table~\ref{table:results} we require values for the rest of the branching
fractions in each product. For the $D^*$  branching fractions
we use the CLEO II results \cite{dstar},
\begin{equation} \begin{array}{llll}
{\cal B}(D^{*+} \rightarrow D^0\pi^+) &= & [68.1 \pm1.0\pm1.3]\%  \\
{\cal B}(D^{*0} \rightarrow D^0\pi^0) &= & [63.6 \pm2.3\pm3.3]\%
\end{array} \label{eqn:dstarbr} \end{equation}
For the $D^0$ branching fraction we use the CLEO II result \cite{kpi}:
\begin{equation} \begin{array}{llll}
{\cal B}(D^0 \rightarrow K^-\pi^+) & = & [3.91 \pm 0.08 \pm 0.07 \pm0.16]\%
\end{array} \label{eqn:dnoradbr} \end{equation}
which is the value given in Ref.\ \cite{kpi} without final state
radiation.  The systematic error on $K$ and $\pi$
track reconstruction (third error) has been separated from other
systematics (second error) because it is correlated with row 5 of
Table~\ref{table:efficiency}. This correlation is an
important advantage of using this CLEO II measurement of
${\cal B}(D^0 \rightarrow K^-\pi^+)$, as  the error due to
$K\pi$ track reconstruction will cancel in the determination of the
$\bar{B} \rightarrow D^*~\ell~\bar{\nu}$ branching fractions.
Thus the fractional
systematic error on our results for
${\cal B}(\bar{B} \rightarrow D^*~\ell~\bar{\nu})$
will be reduced relative to the
error on the product branching fractions listed in Table~\ref{table:results}.

If we assume equal production of charged and neutral meson pairs at the
$\Upsilon(4S)$, $f_{+-} = f_{00} = 0.5$, we obtain,
\begin{eqnarray}
{\cal B}_{0.5}(\bar{B}^0 \rightarrow D^{*+}~\ell^-~\bar{\nu})  & = &
[4.49 \pm 0.32 \pm 0.39]\%, \label{eqn:larry}\\
{\cal B}_{0.5}(B^- \rightarrow D^{*0}~\ell^-~\bar{\nu})  & = &
[5.13 \pm 0.54 \pm 0.64]\%, \label{eqn:curly}
\end{eqnarray}
where the first error is statistical, and the second (systematic)
error includes the errors on the $D^0$ and $D^*$ branching fractions.
On the other hand, by assuming
$\Gamma(\bar{B}^0 \rightarrow D^{*+}~\ell^-~\bar{\nu}) =
\Gamma(B^- \rightarrow D^{*0}~\ell^-~\bar{\nu})$,
which follows from isospin invariance,
and $f_{+-} + f_{00} = 1$, Eqs.\ (\ref{eqn:nzero}) and
(\ref{eqn:nminus}) can be combined to give
\begin{equation}
 \Gamma(\bar{B} \rightarrow D^*~\ell~\bar{\nu})
  =  \frac{1}{4N_{\Upsilon(4S)}{\cal B}_{D^0}} \left[
 \frac{N_0}{\tau_{\bar{B}^0}{\cal B}_{D^{*+}}} +
 \frac{N_-}{\tau_{B^-}{\cal B}_{D^{*0}}} \right].
\label{eq:mgsgamma}
\end{equation}
Using the average of LEP and CDF measurements for the
lifetimes $\tau_{B^-}$ and $\tau_{B^0}$~\cite{lifetimes}
yields,
\begin{equation}
 \Gamma(\bar{B} \rightarrow D^*~\ell~\bar{\nu}) =
[29.9 \pm 1.9 \pm 2.7 \pm 2.0]~ {\rm ns}^{-1},
\label{eqn:pwidth} \end{equation}
independent of $f_{+-}/f_{00}$.  The second (systematic) error
takes into account correlations between the errors of Eqs.\ (\ref{eqn:larry})
and (\ref{eqn:curly}). The third error is due to the $B$ lifetime
measurements and is determined with the conservative assumption
that the errors
on the charged and neutral lifetimes are fully correlated.

The partial width can be converted to either the charged or the
neutral $\bar{B} \rightarrow D^*~\ell~\bar{\nu}$
branching fraction by multiplying Eq.\ (\ref{eq:mgsgamma})
times the appropriate lifetime, but note that
${\cal B}(B^- \rightarrow D^{*0}~\ell^-~\bar{\nu})$ and
${\cal B}(\bar{B}^0 \rightarrow D^{*+}~\ell^-~\bar{\nu})$
are not measured independently this way. The result is
\begin{equation}
{\cal B}(B^- \rightarrow D^{*0}~\ell^-~\bar{\nu}) = \frac{1}{2} \left[
\frac{\tau_{B^-}}{\tau_{\bar{B}^0}}
{\cal B}_{0.5}(\bar{B}^0 \rightarrow D^{*+}~\ell^-~\bar{\nu}) +
{\cal B}_{0.5}(B^- \rightarrow D^{*0}~\ell^-~\bar{\nu}) \right],
\label{eqn:mgsbr} \end{equation}
independent of $f_{+-}/f_{00}$. Note that this does not depend on
the individual $B$ lifetimes, but only on their ratio. Taking
$\tau_{B^-}/\tau_{\bar{B}^0}=1.1\pm0.11$ from other
experiments~\cite{lifetimes} leads to
\begin{equation}
{\cal B}(B^- \rightarrow D^{*0}~\ell^-~\bar{\nu}) =
1.1 \times {\cal B}(\bar{B}^0 \rightarrow D^{*+}~\ell^-~\bar{\nu}) =
[5.03 \pm 0.32 \pm 0.45 \pm 0.24]\%, \label{eqn:allthefuss}
\end{equation}
where the uncertainty in our knowledge of the lifetime ratio leads to the third
systematic error. Use of a different central value for the lifetime
ratio will change the right hand side of Eq.\ (\ref{eqn:allthefuss}) according
to Eq.\ (\ref{eqn:mgsbr}). The factor $1.1$ carries no error to indicate that
we have assumed that the ratio of branching fractions exactly equals the
lifetime ratio, which means that we can not independently determine the
charged and neutral branching fractions.

One can compare the above branching fractions to previous results
from ARGUS and CLEO, which assume $f_{+-} = f_{00} = 0.5$, after correcting all
results for  the new CLEO II $D^*$ and $D^0$ branching fractions \cite{k3p}.
The results of
this comparison are shown in Table~\ref{table:br}.
Our results are in good agreement with previous measurements.

\subsection{$\Upsilon(4S)$ Branching Fractions}

By taking the ratio of Eq.\ (\ref{eqn:nminus}) over (\ref{eqn:nzero})
and assuming only that
$\Gamma(\bar{B}^0 \rightarrow D^{*+}~\ell^-~\bar{\nu}) =
\Gamma(B^- \rightarrow D^{*0}~\ell^-~\bar{\nu})$,
we have
\begin{equation}
 {f_{+-} \over f_{00} } { \tau_{B^-} \over \tau_{\bar{B}^0} } =
{ N_- \over N_0 }
 { {\cal B}_{D^{*+}} \over {\cal B}_{D^{*0}}  } = 1.14 \pm 0.14 \pm 0.13.
\end{equation}
where the first error is statistical and the second (systematic) error
is dominated by the uncertainty on the
slow pion efficiency ratio, $\epsilon_{\pi^0} / \epsilon_{\pi^+}$.
Substituting the value for $\tau_{B^-} / \tau_{\bar{B}^0}$ from
Ref.\ \cite{lifetimes},
\begin{equation}
 {f_{+-} \over f_{00} } = 1.04 \pm 0.13 \pm 0.12 \pm 0.10,
\end{equation}
where the third error is due to the input lifetime ratio.
Finally, the assumption $f_{+-} + f_{00} = 1$ can be used to
extract a value for $f_{+-}$ or $f_{00}$,
\begin{equation}
       f_{+-}  = 1-f_{00} = 0.510 \pm 0.052,
\end{equation}
where the error is  combined statistical and systematic, including the error
in the $B$ lifetime ratio.

\subsection{${\cal B}(\bar{B} \rightarrow
D^*~X~\ell~\bar{\nu})$ from Correlated Background Region Yields}
\label{section:Dtwostar}

The product branching fractions for
${\cal B}(\bar{B} \rightarrow D^*~X~\ell~\bar{\nu})$ in
Table~\ref{table:results} have been computed from the correlated
background yields in the Correlated Background Region
from Tables~\ref{table:numbers0} and
\ref{table:numbers-} divided by an efficiency estimated with Monte Carlo
$\bar{B} \rightarrow D^{**}~\ell~\bar{\nu}$ events, as explained below.

All possible resonant and non resonant sources
of correlated background contribute to the data yields:
\begin{eqnarray}
  N_c(\bar{B} \rightarrow D^*~X~\ell~\bar{\nu}) &=&
{\cal N}\sum_i {\cal B}(\bar{B} \to D^{**}_i~\ell~\bar{\nu})
                   {\cal B}(D^{**}_i \to D^*X) \epsilon_i \nonumber \\
    &+& {\cal N}\sum_j {\cal B}(\bar{B} \to D^*X_j~\ell~\bar{\nu}) \epsilon_j,
\end{eqnarray}
where $i$ ranges over all possible resonant states and $j$ over
all non resonant states, and $\cal N$ contains the total number of
$B$ decays in the data times the $D^*$ and $D^0$ branching fractions.
The efficiencies for the individual channels, $\epsilon_i$ and
$\epsilon_j$, are potentially all different. Therefore, an accurate estimate
of the the efficiency for the
$\bar{B} \rightarrow D^*~X~\ell~\bar{\nu}$ inclusive process,
$\epsilon_c(\bar{B} \rightarrow D^*~X~\ell~\bar{\nu})$, requires
knowledge of the relative abundances of all the resonant and non resonant
exclusive modes. However, if the different exclusive modes all have similar
efficiencies, one need only simulate one or a few of them in order
to estimate $\epsilon_c(\bar{B} \rightarrow D^*~X~\ell~\bar{\nu})$.
We compared the Monte Carlo efficiencies
for $\bar{B} \rightarrow D^{**}~\ell~\bar{\nu}$ where the
$D^{**}$ was a  $1^1P_1$, $1^3P_1$ or $1^3P_2$ state
generated according to the ISGW  model~\cite{isgw} in the $D^*\pi$ decay
channel. Since the $1^3P_1$ resonance is very wide, we expect it to resemble
non resonant decays where $X_j$ is a pion. For these three exclusive channels
we find the following efficiencies in the Correlated Background Region:
$(3.4\pm0.2)\% (1^1P_1)$, $(2.3\pm0.2)\% (1^3P_1)$, and
$(2.9\pm0.2)\%(1^3P_2)$
for a $D^{*+}$ in the final state, and
$(2.5\pm0.2)\%(1^1P_1)$, $(1.7\pm0.2)\%(1^3P_1)$, and $(2.3\pm0.2)\%(1^3P_2)$
for a $D^{*0}$ in the final state. We take the average of
these numbers as a rough estimate of the
Correlated Background Region efficiency for any channel which
leads to a $D^*$ in the final state,
\begin{eqnarray}
\epsilon_c(\bar{B} \rightarrow D^{*+}~X~\ell^-~\bar{\nu})& =& [2.9 \pm0.6]\% \\
\epsilon_c(\bar{B} \rightarrow D^{*0}~X~\ell^-~\bar{\nu})& =& [2.2 \pm0.4]\%,
\end{eqnarray}
where the errors were estimated from the
spread in the efficiencies of the three
individual modes considered, together with the uncertainties from
Table~\ref{table:efficiency}.

Dividing rows 3 and 4 of Table~\ref{table:results} by the
$D^0$ and $D^*$ branching fractions, we obtain,
\begin{eqnarray}
   {\cal B}(\bar{B} \rightarrow D^{*+}~X~\ell^-~\bar{\nu})   & = &
[0.6 \pm 0.3 \pm 0.1]\% \label{eqn:beavis} \\
   {\cal B}(\bar{B} \rightarrow D^{*0}~X~\ell^-~\bar{\nu})   & = &
[0.6 \pm 0.6 \pm 0.1]\% \label{eqn:butthead}
\end{eqnarray}
These inclusive branching fractions do not depend on the unknown relative
abundances of the different resonant and non resonant $D^*X$ states in
semileptonic $B$ decay, as long as they all have similar efficiencies in
the Correlated Background Region. To obtain an upper limit for
${\cal B}(\bar{B} \to D^{**}\ell\bar{\nu})$, we assume that non resonant
channels are all zero, and use the relative abundances of the first radial
excitation $D^{**}$ states in the ISGW model to estimate
${\cal B}(D^{**} \rightarrow D^* \pi) = 77\%$~\cite{scora}.
Together with results (\ref{eqn:beavis}) and (\ref{eqn:butthead}) this leads
to $\sum_i {\cal B}(\bar{B} \to D^{**}_i\ell\bar{\nu}) =
(1.5 \pm0.8 \pm0.2)\%$,
which does not
include an error on the ISGW estimate of the relative $D^{**}$ abundances.
Converting this to an upper limit yields,
\begin{equation}
       \sum_i {\cal B}(\bar{B} \to D^{**}_i\ell\bar{\nu}) < 2.8\%
\end{equation}
at the 95\% confidence level. This result is consistent with previous
model-dependent determinations of the
$\bar{B} \rightarrow D^{**}~\ell~\bar{\nu}$ branching
fraction~\cite{oldargusIII}.

\section{Determination of $|V_{cb}|$}

The decay amplitude of
$\bar{B} \rightarrow D^*~\ell~\bar{\nu}$ for massless leptons \cite{nomass}
is commonly
expressed in terms of the three meson
form factors $A_1(q^2)$, $A_2(q^2)$ and $V(q^2)$
\cite{bsw}. These form factors characterize the transition between the two
strongly bound states $B$ and $D^*$, and are therefore not calculable.
In order to calculate the decay rate for
$\bar{B} \rightarrow D^*~\ell~\bar{\nu}$
and hence determine $|V_{cb}|$
from the measured rate it is necessary to know the
normalization as well as the $q^2$ dependence of these form factors. Several
phenomenological models have been devised in order to estimate the form
factors, but these typically assume some $q^2$ dependence and use ad hoc wave
function representations for the $B$ and $D^*$ bound states. Hence it is
difficult to estimate the accuracy of model predictions. In order to compare
different models we define the form factor ratios,
$A_2/A_1$ and $V/A_1$.  Table ~\ref{table:modused}
lists the expressions for $A_1$, $A_2/A_1$, and $V/A_1$
for the different  models examined.

Recent work on Heavy Quark Effective Theory
and its application to the decay
$\bar{B} \rightarrow D^*~\ell~\bar{\nu}$ has led to constraints
on the form factors that permit
a less model dependent determination of $|V_{cb}|$.  In the limit of
infinite $b$ and $c$ quark masses the
meson form factors $A_1$, $A_2$ and $V$ are well defined functions
of a single form factor $\xi(y)$, known as the Isgur-Wise
function \cite{wisger},
where $y$ is the kinematic variable of HQET, and is related to $q^2$ by
Eq.\ \ref{eqn:three}. The Isgur-Wise function
$\xi(y)$ contains the non perturbative QCD dynamics of the light degrees
of freedom in the mesons, and it is therefore not calculable.
However, at $y=1$ (point of maximum $q^2$) it is normalized to unity.

For finite $b$ and $c$ quark masses the
$\bar{B} \rightarrow D^*~\ell~\bar{\nu}$ differential decay
rate can be expressed in terms of a single unknown form factor
${\cal F}(y)$ which incorporates the three form factors $A_1$, $A_2$, and $V$.
The relationships between these and the Isgur-Wise function are given
by HQET only to leading order, and the corrections necessary to account
for the deviation from the heavy quark symmetry limit are absorbed into
${\cal F}(y)$. Following Neubert~\cite{neubert,etaa} we write,
\begin{eqnarray}
\frac {d{\Gamma}} {dy} & = &
\frac{G^2_F}{48\pi^3}m^3_{D^*}(m_B-m_D^*)^2
 |V_{cb}|^2 {\cal F}^2(y) \times \nonumber \\
&& ~\sqrt{y^2-1}
\biggl\lbrack 4y(y+1)\frac{1-2yr+r^2}{(1-r)^2}+(y+1)^2\biggr\rbrack,
\label{eq:neurate}
\end{eqnarray}
where $r=m_{D^*}/m_B$.
The function ${\cal F}(y)$ can be related to the
Isgur-Wise function and correction terms
that vanish in the infinite $b$ and $c$ mass limit,
$
{\cal F}(1) = \eta_A\xi(1) + {\cal O}\left({\Lambda_{QCD} \over m_Q}\right)^2,
$
where $\eta_A$ is a perturbatively calculable QCD radiative correction and
$\Lambda_{QCD}$ is the QCD scale parameter.
A next to leading order calculation
gives $\eta_A=0.986\pm0.006$~\cite{neubert}.
For $\bar{B} \rightarrow D^*~\ell~\bar{\nu}$ decay,
it has been shown that corrections of order
$1/m_c$ and $1/m_b$ are identically
zero~\cite{luke}, and it is currently estimated
that the second order deviations from
the symmetry limit can be calculated with approximately $4\%$
uncertainty~\cite{ffcorr,mannel,neubert}.
Therefore, a precise measurement of $|V_{cb}|{\cal F}(1)$ will result in
an accurate determination of $|V_{cb}|$ with no need for predictions
of the shape of the decay form factors. Measurements of the differential
decay rate $d\Gamma /dy$ will also determine the shape of ${\cal F}(y)$, which
provides information about the non perturbative QCD dynamics of the decay.

In the next section we
use model predictions for the $A_1$, $A_2$ and $V$ form factors
to extract model dependent values of $|V_{cb}|$ from our measurement of
${\cal B}(\bar{B}^0 \rightarrow D^{*+}~\ell^-~\bar{\nu})$ and
${\cal B}(B^- \rightarrow D^{*0}~\ell^-~\bar{\nu})$.
These model dependent extractions have unknown
theoretical errors. Following the model dependent extractions we
describe the measurement of $|V_{cb}|{\cal F}(1)$ from the differential
decay distribution $d\Gamma(\bar{B} \rightarrow D^*~\ell~\bar{\nu})/dy$.
Using a model
prediction for the corrections to the heavy quark
symmetry limit at $y=1$ we then present a value for $|V_{cb}|$.
We also compare the results for the shape of ${\cal F}(y)$ with
model predictions.

\subsection{Model dependent determinations}

There are several quark models that
predict the normalization and $q^2$ dependence of the
form factors describing the decay $\bar{B} \rightarrow D^*~\ell~\bar{\nu}$.
When extracting model dependent values for $|V_{cb}|$
based on predictions of the form factors and their
$q^2$ dependence it is important to determine the efficiency for the model
under investigation because the acceptance can vary with different
values of $A_1$, $A_2/A_1$ and $V/A_1$.
The form factor ratios and their $q^2$ dependence for the
different models considered here are in Table~\ref{table:modused}.
Using the partial width from Eq.\ (\ref{eqn:pwidth}),
$|V_{cb}|$ can be calculated from model predictions of the
rate for $\bar{B} \rightarrow D^*~\ell~\bar{\nu}$.
The calculation includes a correction
for the difference in the efficiency predicted with each model;
the values for $|V_{cb}|$ obtained with four
different models are given in Table~\ref{table:vcbmod}.

\subsection{$|V_{cb}|$ determined from the
$d\Gamma(\bar{B} \rightarrow D^*~\ell~\bar{\nu})/dy$ distribution}

In this section the extraction of $|V_{cb}|{\cal F}(1)$ from the
$d\Gamma(\bar{B} \rightarrow D^*~\ell~\bar{\nu})/dy$
distribution is described.  Three fits are performed.
Using the assumption that $f_{+-}=f_{00}=0.5$, we first extract
$|V_{cb}|{\cal F}(1)$
separately from  the differential
decay distributions for
$\bar{B}^0 \rightarrow D^{*+}~\ell^-~\bar{\nu}$ and for
$B^- \rightarrow D^{*0}~\ell^-~\bar{\nu}$ events.
We then simultaneously fit the
$\bar{B}^0 \rightarrow D^{*+}~\ell^-~\bar{\nu}$ and
the $B^- \rightarrow D^{*0}~\ell^-~\bar{\nu}$ events,  which results  in a
determination of $|V_{cb}|{\cal F}(1)$
that  is independent of $f_{00}/f_{+-}$.

We use the unbinned maximum likelihood
method developed for fitting the multidimensional decay distribution of
$D \to K^* l \bar{\nu}$ events ~\cite{schmidt}.  Application to
the extraction of $|V_{cb}|{\cal F}(y)$ is straightforward since only
one dimension is used in the fit. It is convenient to make the likelihood
a function of $V_{cb}$ by normalizing our probability distribution function to
the number of observed events. This requires the inclusion of a Poisson
probability factor~\cite{cassel}.

$D^*$ and lepton pairs are selected in the Signal Region of the $C$ vs. $MM^2$
plane as already described. However, since we are now performing an unbinned
likelihood fit, we apply a $M_{K\pi}$ cut as well as a $\delta_m$
cut. The $M_{K\pi}$
cuts are $| M_{K\pi} - m_{D^0} | < 25$~MeV  and
$| M_{K\pi} - m_{D^0} | < 20$~MeV
for $\bar{B}^0 \rightarrow D^{*+}~\ell^-~\bar{\nu}$ and
$B^- \rightarrow D^{*0}~\ell^-~\bar{\nu}$ respectively, where
$m_{D^0}$ is the nominal $D^0$ mass.
In order to have low background levels in the unbinned likelihood fit,
the tighter
$B^- \rightarrow D^{*0}~\ell^-~\bar{\nu}$
cut was chosen to optimize signal to background as much as
possible, without introducing a systematic bias from
reproduction of signal shapes by the Monte Carlo simulation.

Eq.\ (\ref{eq:neurate}) can be rewritten as
\begin{equation}
{d\Gamma \over dy} = {\cal G}(y) |V_{cb}|^2{\cal F}^2(y),
\label{eq:dgdy}
\end{equation}
where all the known quantities have been folded into a single
function ${\cal G}(y)$. $|V_{cb}|$ and ${\cal F}(y)$
are to be determined by fitting the data. Since we want to
determine the product $|V_{cb}|{\cal F}(y)$ at $y=1$,
we approximate the
unknown function ${\cal F}(y)$ with an expansion about $y=1$,
\begin{equation}
{\cal F}(y) = {\cal F}(1)\left[1 - a^2(y-1) + b(y-1)^2 \right],
\end{equation}
thus defining the fit parameters to be $|V_{cb}|{\cal F}(1)$, $a^2$ and $b$.
We use $a^2$ rather than $a$ because the $B$ and $D^*$ meson wave functions
have maximum overlap at $y=1$, so the first derivative of ${\cal F}(y)$
must be negative at that point. Our results for $a^2$ and $b$
can be used to compare the shape of ${\cal F}(y)$ determined from data with
theoretical model predictions~\cite{slope}.

The variable $y$ is  the $D^*$ energy to mass ratio in the
rest frame of the decaying $B$. However, in data we measure
$E_{D^*}/m_{D^*}$ in the laboratory frame, symbolized by $\tilde{y}$.
This quantity is a function of $y$ and the $B$ meson momentum ${\bf p}_B$,
but since the direction of ${\bf p}_B$ is unknown, the $dN/dy$
distribution of the data is not directly observed.
We therefore fit $dN/d{\tilde y}$, and naturally incorporate
the smearing due to the $B$ motion in the same way as
the smearing due to detector resolution (which is much smaller).
This is accomplished by a function ${\cal R}(\tilde{y},y)$, which
is convoluted with ${\cal G}(y){\cal F}(y)$ in order to arrive at a
probability distribution function that can be compared with the data,
\begin{equation}
   G(\tilde{y},a,b)=
     \int_1^{y_0}{\cal G}(y){\cal F}^2(y){\cal R}(y,\tilde{y})dy,
\end{equation}
where $y_0$ is the upper kinematic limit of $y$, given by $q^2=0$.
The resolution function ${\cal R}(\tilde{y},y)$ is determined from
Monte Carlo simulation, and allows $G(\tilde{y},a,b)$
to correctly include changes in detection efficiency with variations
in the parameters of ${\cal F}$, thus reducing model dependence.

\begin{figure}
\centerline{\psboxto(6.0in;6.0in){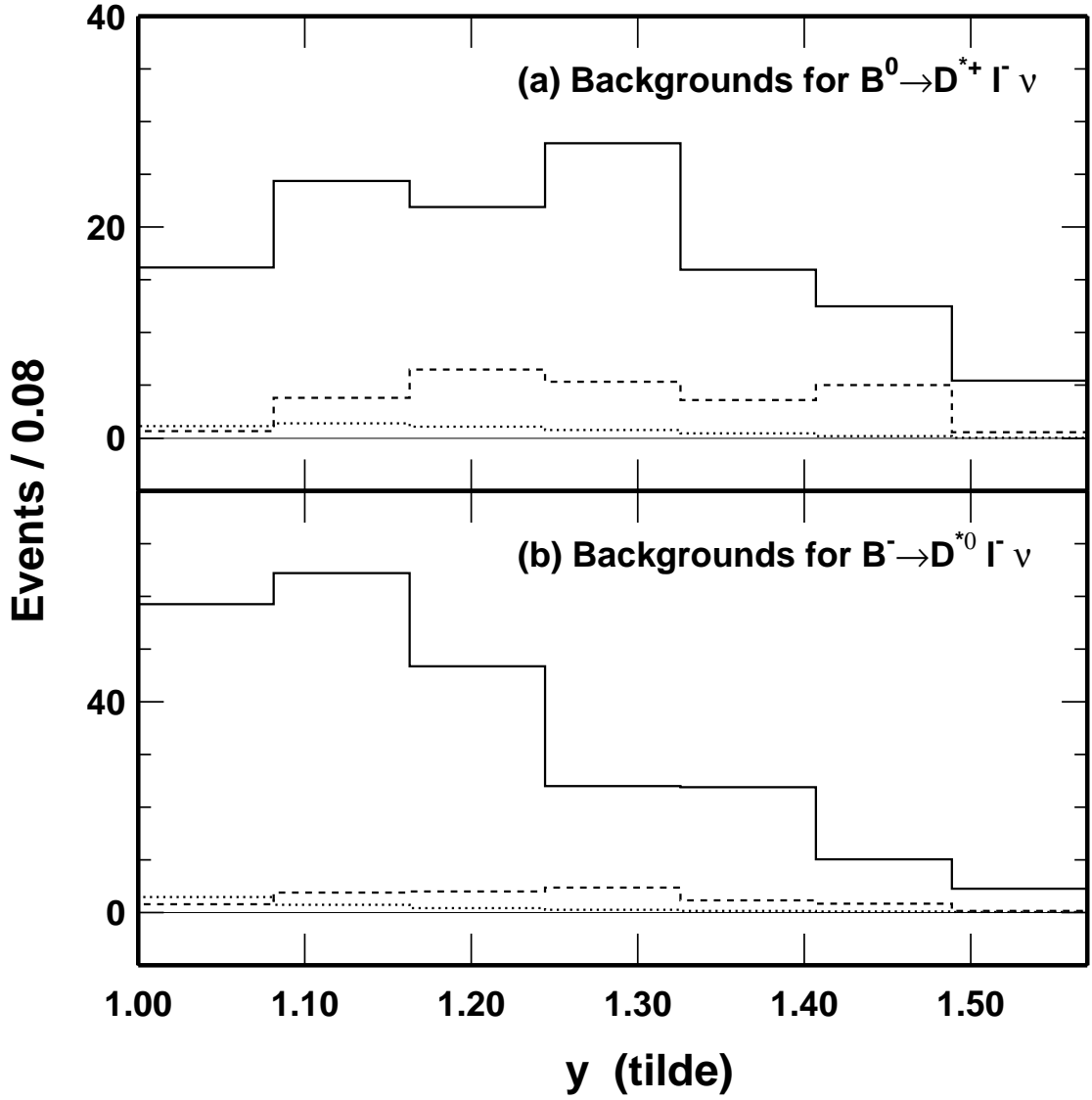}}
 \caption{The $\tilde{y}$ distributions of the
backgrounds for (a) $\bar{B}^0 \rightarrow D^{*+}~\ell^-~\bar{\nu}$ events
and for (b) $B^- \rightarrow D^{*0}~\ell^-~\bar{\nu}$ events.
The solid line is the combinatoric
background, the dashed line is the correlated background, and
the dotted line is the uncorrelated background.  The area of the
curves are normalized to represent
the background levels in the data.}
  \label{figure:yback}
\end{figure}

The full likelihood function is
\begin{eqnarray}
{\cal L}(|V_{cb}|,a,b)&=&
e^{-N(|V_{cb}|,a,b)-N_b}
\prod_{i=1}^{n}\left(
{N(|V_{cb}|,a,b)G(\tilde{y}_i,a,b)+
\sum_{b}n_{b}P_b(\tilde{y_i})}\right),
\label{eq:likelihood}
\end{eqnarray}
with
\begin{eqnarray}
N(|V_{cb}|,a,b) &=&
4N_{\Upsilon(4S)}f{\cal B}_{D^*}{\cal B}_{D^0}\tau_B|V_{cb}|^2\int_{1}^{y_0}
\epsilon(y){\cal G}(y){\cal F}^2(y) dy
\label{eq:norm}  \\
N_b &=& n_{{\rm comb}}+n_{D^{**}l\bar{\nu}}+n_{{\rm uncorr}}
\label{eq:backnorm}  \\
\sum_b n_b P_b(\tilde{y_i})
                 &=&   n_{{\rm comb}}P_{{\rm comb}}(\tilde{y_i})
                     + n_{D^{**}l\bar{\nu}}P_{D^{**}l\bar{\nu}}(\tilde{y_i})
                     + n_{{\rm uncorr}}P_{{\rm uncorr}}(\tilde{y_i}).
\label{eq:backlike}
\end{eqnarray}
$N(|V_{cb}|,a,b)$ is the expected number of reconstructed signal events, where
$\epsilon(y)$ is the efficiency for detecting
$\bar{B} \rightarrow D^*~\ell~\bar{\nu}$ events determined
from the Monte Carlo, $\tau_B$ is the $B$ meson lifetime and
$f$ (meaning either $f_{+-}$ or $f_{00}$), $N_{\Upsilon(4S)}$,
${\cal B}_{D^*}$ and ${\cal B}_{D^0}$ have been defined in
Eqs.\ \ref{eqn:nzero} and \ref{eqn:nminus}.
The total number of background events passing our cuts is $N_b$.
The exponential factor in Eq.\ (\ref{eq:likelihood})
comes from the Poisson probability mentioned earlier.
The backgrounds are  represented by Eqs.\ (\ref{eq:backnorm}) and
(\ref{eq:backlike}).  They
have been divided into three types, combinatoric,
$\bar{B} \rightarrow D^*~X~\ell~\bar{\nu}$, and
uncorrelated background,  and their normalization and $\tilde{y}$
dependence is represented by Eq.\ (\ref{eq:backlike}).
The combinatoric background yield and $\tilde{y}$ distribution,
($n_{{\rm comb}}P_{{\rm comb}}(\tilde{y})$),
are obtained from sidebands in the $\delta m$
distribution.  The correlated background
yield, $n_{D^{**}l\bar{\nu}}$,
is obtained from Table~\ref{table:numbers0} or
{}~\ref{table:numbers-}~\cite{newcuts}
and its $\tilde{y}$ distribution,  $ P_{D^{**}l\bar{\nu}}(\tilde{y_i})$,
is taken from Monte Carlo generated
as in Section \ref{section:Dtwostar} according to the ISGW model. The
uncorrelated background yield and $\tilde{y}$ distribution,
$n_{{\rm uncorr}}P_{{\rm uncorr}}(\tilde{y})$
is obtained from the measured inclusive $B \to D^*$ and
$B\to l^-$ spectra, assuming an isotropic
angular distribution  between the uncorrelated leptons and $D^*$'s.
The three $\tilde{y}$ distributions of the backgrounds
to $\bar{B}^0 \rightarrow D^{*+}~\ell^-~\bar{\nu}$ and
$B^- \rightarrow D^{*0}~\ell^-~\bar{\nu}$
are shown in Fig.\ \ref{figure:yback}.

\begin{figure}
\centerline{\psboxto(6.0in;6.0in){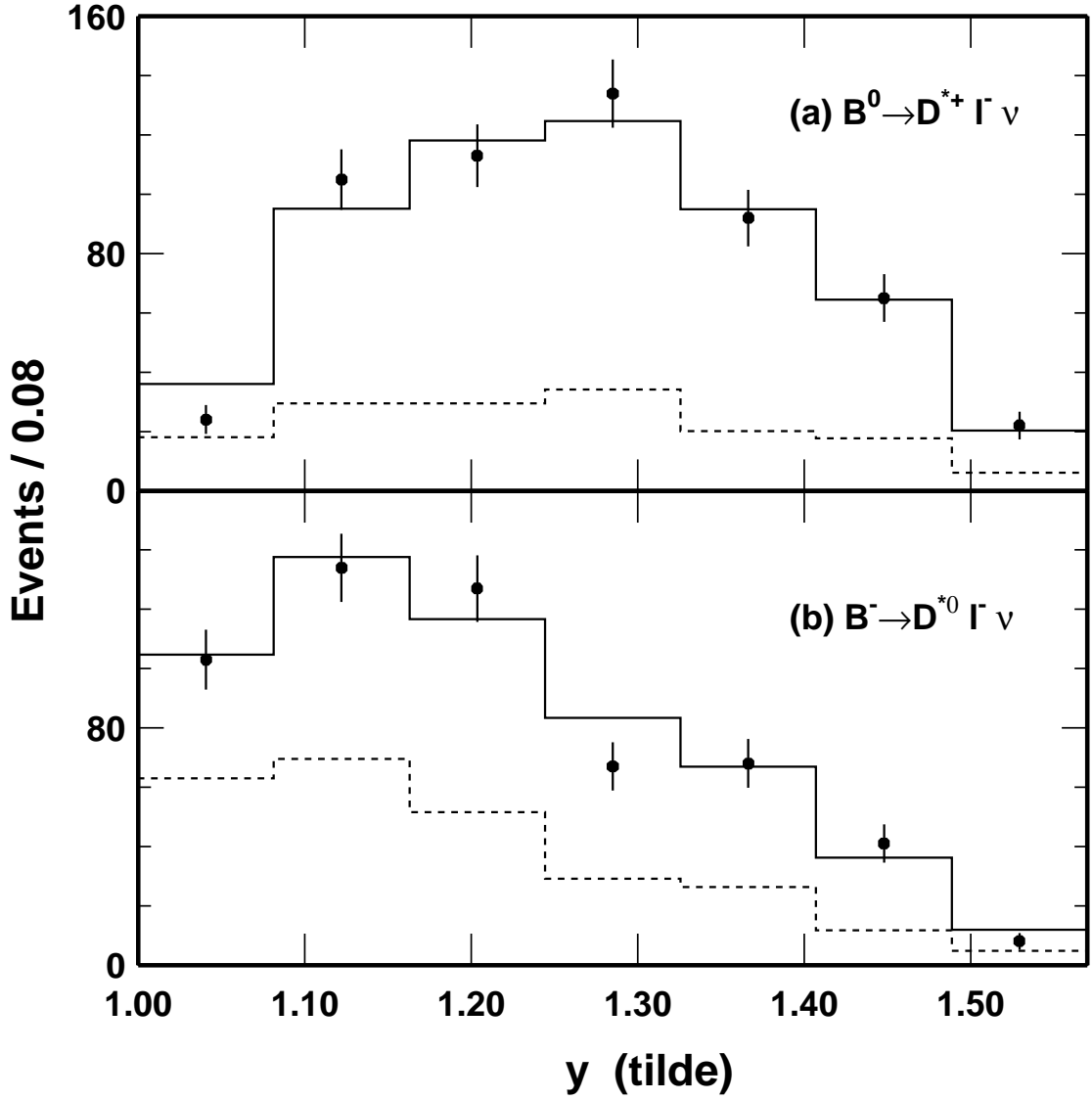}}
 \caption{Differential yield, $dN/d\tilde{y}$, for (a)
$\bar{B}^0 \rightarrow D^{*+}~\ell^-~\bar{\nu}$ events
and (b) $B^- \rightarrow D^{*0}~\ell^-~\bar{\nu}$ events
in the data, with the projections of the unbinned fits superimposed.
The solid histogram represents the result of the fit using
${\cal F}(y) = 1 - a^2(y-1)$.
The dashed histogram shows the level of the background from all sources.}
  \label{figure:dg/dy}
\end{figure}

The data yields as a function of $\tilde{y}$ for
$\bar{B}^0 \rightarrow D^{*+}~\ell^-~\bar{\nu}$ and
$B^- \rightarrow D^{*0}~\ell^-~\bar{\nu}$ events are shown  in
Fig.\ \ref{figure:dg/dy}.
The values for $|V_{cb}|{\cal F}(1)$ and $a^2$ from linear
fits to these data are given in Table~\ref{table:fitvcb}.
We have fixed $b$ to zero in these fits because
the statistics of the individual modes are not sufficient
to constrain the curvature of ${\cal F}(y)$.
The first error listed in Table~\ref{table:fitvcb} is
statistical and the second error is an estimate of the systematic
error, including the uncertainty in the $B$ lifetimes.
The statistical uncertainty in the backgrounds
has been taken into account by varying the background levels in the fit
by $\pm 1$ standard deviation of their measured central values,
and is included in the statistical error.

\begin{figure}
\centerline{\psboxto(6.0in;6.0in){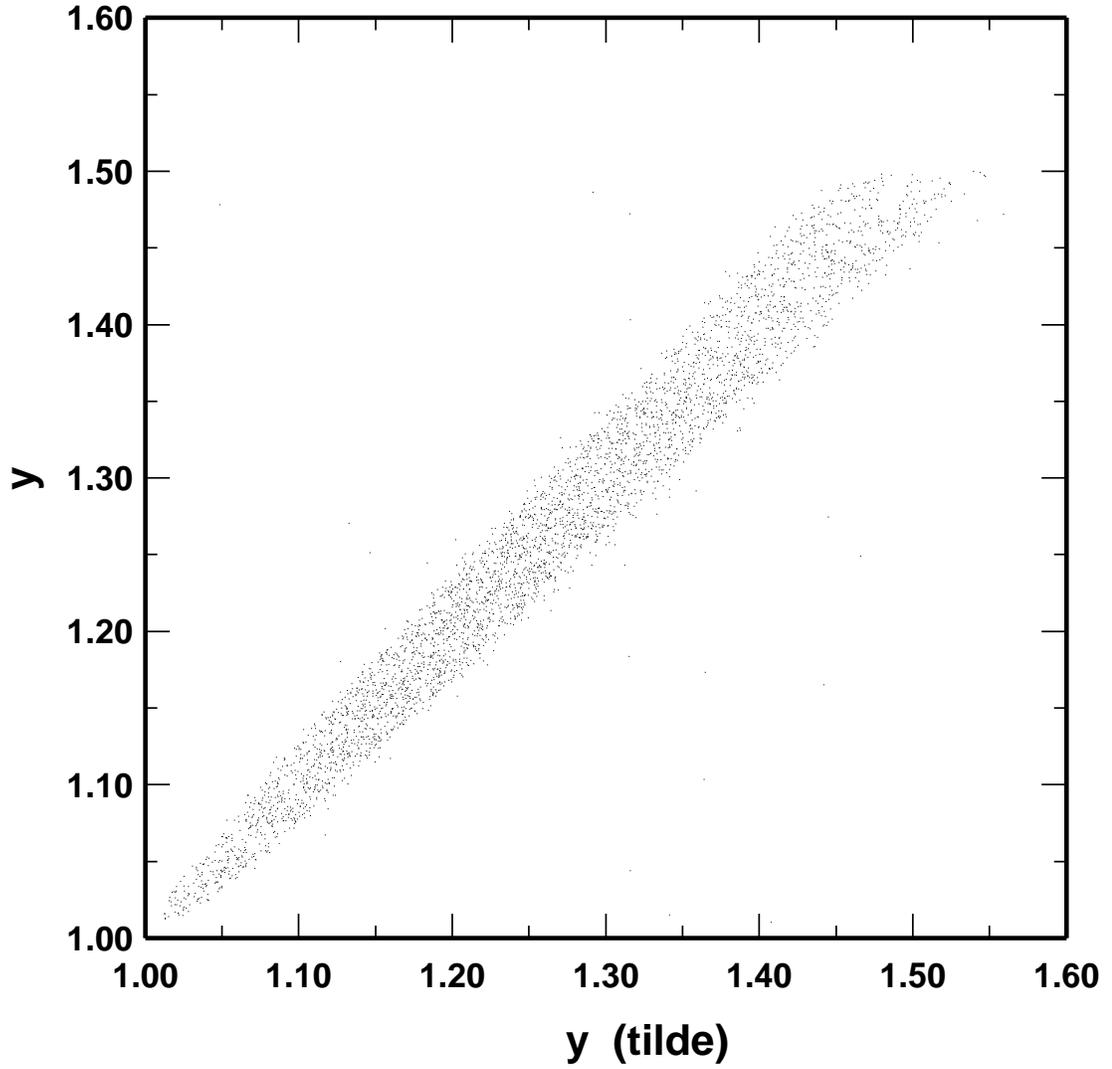}}
 \caption{The smearing between the variables $y$ and $\tilde{y}$ for a
lepton in the momentum  range 1.4 to 2.4~GeV. The points are from Monte Carlo
events generated according to the model of Ref.\ \protect{\cite{neubert}}.}
  \label{fig:yvsytilde}
\end{figure}

The contributions to the systematic error on $|V_{cb}|{\cal F}(1)$ and $a^2$
are listed in Table~\ref{table:sysvcb}.
The errors on the absolute efficiencies are the same as for the
branching fraction measurements.  The errors that are detailed in
Table~\ref{table:sysvcb} are specific to the $d\Gamma/dy$ fits
and account for the uncertainty in the efficiency as a function of
$\tilde{y}$ and uncertainties in the fitting procedure.
As a check of the fitting technique we have performed a binned fit to
the efficiency corrected $d\Gamma/d\tilde{y}$ distribution, using
an analytically smeared form~\cite{mgsthesis},
\begin{equation}
 G'(\tilde{y},a,b) = {1 \over 2 \gamma\beta}
\int_{y_{min}(\tilde{y})}^{y_{max}(\tilde{y})}
{ {\cal G}(y) {\cal F}^2(y) \over \sqrt{y^2-1} } dy,
\end{equation}
where $\gamma\beta$ is the known momentum to mass ratio of the decaying
$B$ meson and ($y_{min}$,$y_{max}$) is the range of possible
values of $y$ given a value of $\tilde{y}$ (see Fig. \ref{fig:yvsytilde}).
The results agree well with the results
obtained from the unbinned likelihood fit.
The systematic uncertainties due to the fitting procedure were estimated
by comparing the results of the unbinned and binned methods in 36
Monte Carlo samples, each with the same statistics as our data sample.

\begin{figure}
\centerline{\psboxto(6.0in;6.0in){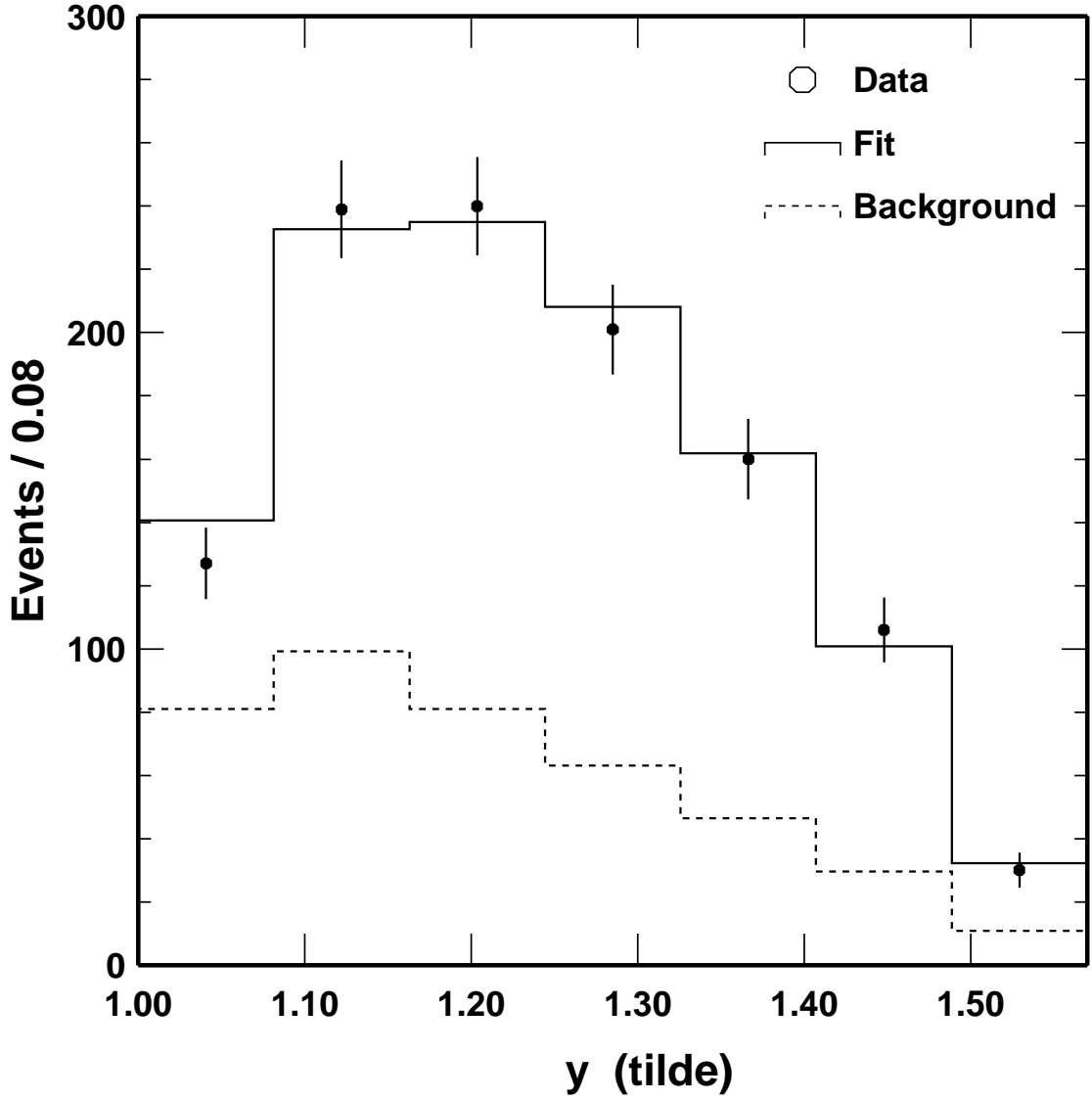}}
 \caption{Differential yield, $dN/d\tilde{y}$, for
$B^- \rightarrow D^{*0}~\ell^-~\bar{\nu}$ events  and
$\bar{B}^0 \rightarrow D^{*+}~\ell^-~\bar{\nu}$ events
combined with the projection of the unbinned simultaneous fit superimposed.
The solid histogram represents the result of the fit using
${\cal F}(y) = 1 - a^2(y-1)$.
The dashed histogram shows the sum of the background levels.}
  \label{fig:comvcb}
\end{figure}

We have also fit the differential decay distribution of the two decays
modes $\bar{B}^0 \rightarrow D^{*+}~\ell^-~\bar{\nu}$ and
$B^- \rightarrow D^{*0}~\ell^-~\bar{\nu}$ simultaneously by calculating the
likelihood of the two modes separately and maximizing their product.
The result of the simultaneous fit is shown in Fig.~\ref{fig:comvcb},
along with the combined data and  background level
for the $\bar{B}^0 \rightarrow D^{*+}~\ell^-~\bar{\nu}$ and
$B^- \rightarrow D^{*0}~\ell^-~\bar{\nu}$ modes.
{}From a linear fit to these data we obtain
the most precise measurement to date of
$|V_{cb}|{\cal F}(1)$ and $a^2$~\cite{correlation},
\begin{eqnarray}
|V_{cb}|{\cal F}(1) & = &
0.0351 \pm 0.0019 \pm 0.0018 \pm 0.0008 \label{eqn:vcbfbds} \\
          a^2  & = & 0.84 \pm 0.12 \pm 0.08,
\label{eqn:asquared} \end{eqnarray}
independent of $f_{+-}/f_{00}$; where the first error is statistical, the
second is systematic, and the
third is due to the input lifetimes~\cite{liferror}.
With the statistics of both modes combined we can
also perform a three parameter fit, where the quadratic term in the
expansion for ${\cal F}(y)$ is retained. The results of this fit
are shown in the last row of Table~\ref{table:fitvcb}.
For comparison, we provide in Table~\ref{table:abmodel}
the $a^2$ (and $b$) values predicted by the different decay models.

To extract a value of $|V_{cb}|$ one requires a prediction for
${\cal F}(1)$. Using ${\cal F}(1)=0.97\pm0.04$~\cite{neubert,nineseven}
we obtain,
\begin{equation}
|V_{cb}| = 0.0362 \pm 0.0019 \pm 0.0020 \pm 0.0014,
\label{eqn:whybother} \end{equation}
where the third error is the quoted theoretical uncertainty in
${\cal F}(1)$ and
the uncertainty in the $B$ lifetimes has been included in the
systematic (second) error~\cite{liferror}.
This result is consistent with previously
published values of $|V_{cb}|$ (after scaling for differences in $D$ and $D^*$
branching fractions and $B$ lifetimes)
from both inclusive and exclusive analyses, although the unknown
model dependence of the previous $|V_{cb}|$ extractions makes
detailed comparisons difficult.
Model dependent values of $|V_{cb}|$ in Table~\ref{table:vcbmod} are
also consistent with this result.

Other recent estimates are ${\cal F}(1)=0.96\pm0.03$~\cite{mannel} and
${\cal F}(1) < 0.94$~\cite{shifman}. The value of ${\cal F}(1)$ used to obtain
Eq.\ (\ref{eqn:whybother}) is consistent with both of these results.
However, Ref.\ \cite{shifman} also claims an ``educated guess'' of
${\cal F}(1) = 0.89\pm0.03$, which would lead to
$|V_{cb}| = 0.0395 \pm 0.0021 \pm 0.0022 \pm 0.0012$.

\begin{figure}
\centerline{\psboxto(6.0in;6.0in){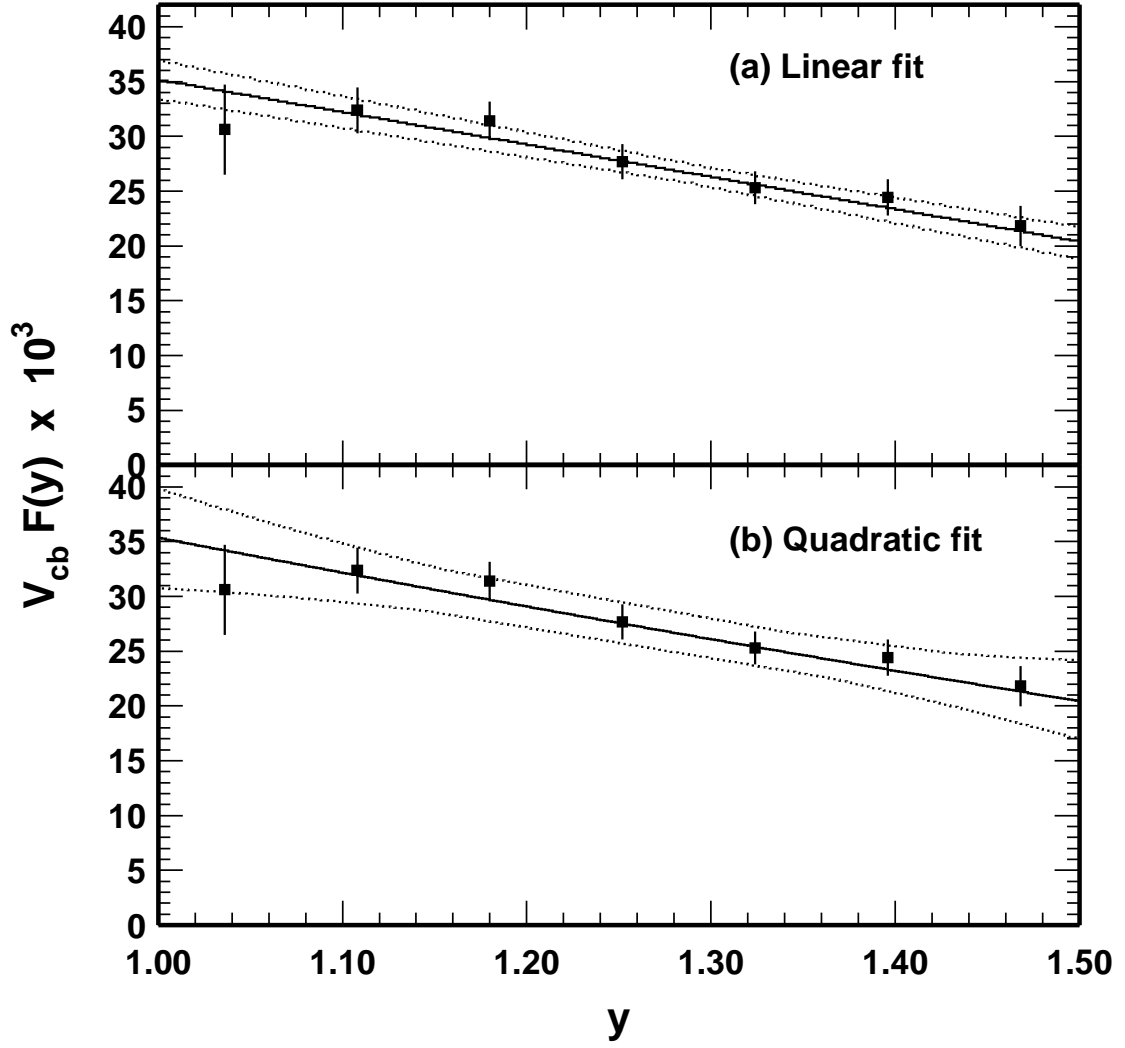}}
 \caption{The product $|V_{cb}|{\cal F}(y)$
(solid line) as determined by our fits
fits to the combined $D^{*+}l^-$ and $D^{*0}l^-$ data,
using (a) a linear expansion and (b) a quadratic expansion
of the form factor ${\cal F}(y)$.
The dotted lines show the contours for $1\sigma$ variations of the
fit parameters. The points are data for the square root of the measured decay
rate divided by the factors other than ${\cal F}(y)$ and $V_{cb}$ of
Eq.\ (\protect{\ref{eq:neurate}}) (error bars are statistical only).
As explained in the text, the data are binned in $y_A(\tilde{y})$, which
is not an unbiased estimator of $y$. Therefore, the data points
do not exactly correspond to the product $|V_{cb}|{\cal F}(y)$.}
  \label{fig:xeofy}
\end{figure}

The product $|V_{cb}|{\cal F}(y)$ as determined from
our fits to the combined $\bar{B}^0 \rightarrow D^{*+}~\ell^-~\bar{\nu}$ and
$B^- \rightarrow D^{*0}~\ell^-~\bar{\nu}$ data
is shown in Fig.\ \ref{fig:xeofy}. Fig.\ \ref{fig:xeofy}(a)
shows the result obtained by keeping only the linear term in the
expansion of ${\cal F}(y)$, while (b) shows the result for a second order
polynomial. The dotted lines show the contours for $1\sigma$ variations of the
fit parameters. All the lines are functions of the variable $y$, which
is equal to the $D^*$ energy to mass ratio, $E_{D^*}/m_{D^*}$, evaluated
in the $B$ rest frame. The points are the data.  As discussed before,
we do not measure $y$ but rather $\tilde{y}$, which is $E_{D^*}/m_{D^*}$
in the laboratory frame. We have therefore binned the data in
the average of the maximum and minimum possible values of $y$ for each
measured $\tilde{y}$. This average is denoted by $y_A(\tilde{y})$.
Unfortunately $y_A(\tilde{y})$ is not an unbiased estimator or $y$, and
although $y_A(\tilde{y})$ has the same kinematic limits as $y$, the
smearing between these two variables is significant compared to the
bin size. Therefore,
the data shown this way do not exactly correspond to the product
$|V_{cb}|{\cal F}(y)$. Note, however, that the lines shown are
not the result of a fit to these data points, but were obtained from our
unbinned fit which correctly accounts for the boost between the $B$ frame and
the laboratory frame.

The shape of ${\cal F}(y)$ determined from the fit to the data
can be used to obtain the functional form of
$d\Gamma(\bar{B} \rightarrow D^*~\ell~\bar{\nu})/dq^2$;
(where $q^2$ is the true $q^2$ of the decay,
not the smeared variable measurable
experimentally) which is useful to test hypotheses of
factorization in $B$ meson decay~\cite{bigb}.
Values of $d\Gamma(\bar{B} \rightarrow D^*~\ell~\bar{\nu})/dq^2$
as determined from the data
are given in Table~\ref{table:dgdqsquared} for commonly used $q^2$ points.

\section{Conclusions}

Using the CLEO II data sample we have reconstructed
both  $\bar{B}^0 \rightarrow D^{*+}~\ell^-~\bar{\nu}$ and
$B^- \rightarrow D^{*0}~\ell^-~\bar{\nu}$ decay modes in the
$D^*$ decay chains:
$D^{*+} \rightarrow D^0\pi^+$ and  $D^{*0} \rightarrow D^0
\pi^0$ with $D^0 \rightarrow K^-\pi^+$.
We have measured the exclusive branching fractions,
\begin{eqnarray}
  {\cal B}(\bar{B}^0 \rightarrow D^{*+}~\ell^-~\bar{\nu}) =
(0.5/f_{00})[4.49 \pm 0.32({\rm stat.}) \pm 0.39({\rm sys.})]\% \\
  {\cal B}(B^- \rightarrow D^{*0}~\ell^-~\bar{\nu}) =
(0.5/f_{+-})[5.13 \pm 0.54({\rm stat.}) \pm 0.64({\rm sys.})]\%,
\end{eqnarray}
which depend on the relative production of charged and
neutral $B$ mesons at the $\Upsilon(4S)$ resonance. We have also
presented these results in a way that requires no knowledge
of $f_{+-}/f_{00}$, but does depend on $B$ lifetime measurements
from other experiments. With the assumption of equal partial widths,
$\Gamma(\bar{B}^0 \rightarrow D^{*+}~\ell^-~\bar{\nu}) =
\Gamma(B^- \rightarrow D^{*0}~\ell^-~\bar{\nu})$,
we have determined the product
\begin{eqnarray}
{f_{+-} \over f_{00}}{\tau_{B^+} \over \tau_{B^0}} =
1.14 \pm0.14{\rm (stat.)} \pm0.13{\rm (sys.)},
\end{eqnarray}
and by using existing $B$ lifetime ratio measurements~\cite{lifetimes}
we have solved for
$f_{+-}/f_{00}=1.04 \pm 0.13{\rm(stat.)}
\pm0.12{\rm (sys.)} \pm0.10{\rm (lifetime)}$.
This measurement verifies  at the $10\%$ level the assumption that
$f_{+-} = f_{00}$, widely used to calculate $B$ meson branching
fractions with data collected at the $\Upsilon(4S)$ resonance.
Also with the assumptions of equal partial widths and $f_{+-}+f_{00}=1$,
and using external
$B$ lifetime measurements, we have determined the partial width
\begin{equation}
 \Gamma(\bar{B} \rightarrow D^*~\ell~\bar{\nu}) =
[29.9 \pm 1.9({\rm stat.}) \pm2.7({\rm sys.})
\pm2.0({\rm lifetime})]{\rm ns}^{-1},
\end{equation}
independent of $f_{+-}/f_{00}$. From this width we have determined
the $\bar{B} \rightarrow D^*~\ell~\bar{\nu}$
branching fractions with dependence only on the
charged to neutral $B$ lifetime ratio (but not the individual lifetimes),
\begin{eqnarray}
{\cal B}(B^- \rightarrow D^{*0}~\ell^-~\bar{\nu}) &=&
1.1 \times {\cal B}(\bar{B}^0 \rightarrow D^{*+}~\ell^-~\bar{\nu}) \nonumber \\
                                                        &=&
[5.03 \pm 0.32{\rm (stat.)} \pm 0.45{\rm (sys.)} \pm 0.24{\rm (lifetime)}]\%.
\end{eqnarray}
where the factor $1.1$ carries no error
because we are assuming that the ratio of branching fractions exactly equals
the value used for the lifetime ratio
(the uncertainty in our knowledge of the lifetime ratio
results in the third error).

Taking advantage of theoretical constraints on the normalization and $q^2$
dependences of the form factors for the
$\bar{B} \rightarrow D^*~\ell~\bar{\nu}$ decay provided by HQET,
a combined fit to the differential decay distributions for
$\bar{B}^0 \rightarrow D^{*+}~\ell^-~\bar{\nu}$ and
$B^- \rightarrow D^{*0}~\ell^-~\bar{\nu}$ results in a
determination of $|V_{cb}|{\cal F}(1)$,
\begin{equation}
 |V_{cb}|{\cal F}(1) =
0.0351 \pm 0.0019{\rm (stat.)}
\pm 0.0018{\rm (sys.)} \pm 0.0008{\rm (lifetime)},
\end{equation}
which is also independent of $f_{+-}/f_{00}$. The explicit
dependence on the $B^+$ and $B^0$
lifetimes is given in Ref.\ \cite{liferror}.
The form factor
normalization ${\cal F}(1)$ is believed to be predictable with small
theoretical uncertainty, and this permits a precise determination of
the CKM matrix element. Using the prediction of Ref.\ \cite{ffcorr,neubert}
for the normalization of ${\cal F}$, we obtain
$|V_{cb}| = 0.0362 \pm 0.0019{\rm (stat.)}
\pm 0.0020{\rm (sys.)} \pm 0.0014{\rm (model)}$, where the
uncertainty in the $B$ lifetimes is included in the systematic error.

\section{Acknowledgements}
We gratefully acknowledge the effort of the CESR staff in providing us with
excellent luminosity and running conditions.
J.P.A. and P.S.D. thank
the PYI program of the NSF, I.P.J.S. thanks the YI program of the NSF,
G.E. thanks the Heisenberg Foundation,
K.K.G., I.P.J.S., and T.S. thank the
%
%
 TNRLC,
K.K.G., H.N.N., J.D.R., T.S.  and H.Y. thank the
OJI program of DOE
and P.R. thanks the A.P. Sloan Foundation for
support.
This work was supported by the National Science Foundation and the
U.S. Dept. of Energy.
\begin{table}
\caption{\label{table:numbers0}
A summary of the events in the
$\bar{B}^0 \rightarrow D^{*+}~\ell^-~\bar{\nu}$ Signal Region
and Correlated Background Region. In the top 2 rows
the first error is statistical and the second is the systematic
uncertainty
due to the choice of functions used to fit the data. The bottom two rows have
been computed using Eq.\ (\protect{\ref{eqn:simultaneous}}) and the
uncertainties in $r_s$ and $r_c$
have been propagated into the systematic error.
The numbers in parenthesis are not independent measurements,
but merely functions
of the other numbers in the bottom two rows.}
\begin{tabular}{ccc}
 $D^{*+}l^-$                  & Signal  & Correlated \\
                              & Region  & Background Region \\
\hline
 $\delta_m$ Signal  & $457 \pm 23 \pm 9 $ & $70 \pm 11 \pm 1 $ \\
 Scaled $\delta_m$ Sideband  & $47 \pm 6 \pm 1 $ & $9 \pm 5$ \\
 Uncorrelated Background & $5.0 \pm 0.5$ & $19 \pm 2$ \\
 Continuum  & $5 \pm 7$  & $0 \pm 6$ \\
 Correlated Lepton Fakes  & $2 \pm 1$ & $1 \pm 1$ \\
 $B \to D^*\tau\bar{\nu}$ and $B \to D^*D_s$ & -- & $4\pm3$ \\
\hline
Net Yield ($N_S$ and $N_C$)  & $398 \pm 25 \pm 9 $ & $37 \pm 14 \pm 1 $ \\
\hline
 $N_s(\bar{B}^0 \rightarrow D^{*+}~\ell^-~\bar{\nu})$ &$376 \pm27 \pm 16$ &
 ($6\pm0.5$) \\
 $N_c(\bar{B}^0 \rightarrow D^{*+}~\ell^-~\bar{\nu})$ & ($22\pm9\pm11$) &
$31 \pm 14 \pm 1$ \\
\end{tabular}
\end{table}

\begin{table}
\caption{ \label{table:numbers-}
A summary of the events in the
$B^- \rightarrow D^{*0}~\ell^-~\bar{\nu}$ Signal Region
and Correlated Background Region. In the top 2 rows
the first error is statistical and the second is the systematic
uncertainty
due to the choice of functions used to fit the data. The bottom two rows have
been computed using Eq.\ (\protect{\ref{eqn:simultaneous}}) and the
uncertainties in $r_s$ and $r_c$
have been propagated into the systematic error.
The numbers in parenthesis are not independent measurements,
but merely functions
of the other numbers in the bottom two rows.}
\begin{tabular}{ccc}
 $D^{*0}l^-$        & Signal  & Correlated \\
                    & Region  & Background Region \\
\hline
 $\delta_m$ Signal  & $476 \pm 25 \pm 10$ & $94 \pm 15 \pm 1$ \\
 Scaled $\delta_m$ Sideband  & $144 \pm 10 \pm 3$ & $34 \pm 7$ \\
 Uncorrelated Background & $8 \pm 1$ & $23 \pm 3$ \\
 Continuum & $6 \pm 7$  & $9 \pm 11$ \\
 Correlated Lepton Fakes & $2 \pm 2$ & $1 \pm 1$ \\
 $B \to D^*\tau\bar{\nu}$ and $B \to D^*D_s$ & - & $2\pm2$ \\
\hline
Net Yield ($N_S$ and $N_C$)   & $ 316 \pm 28 \pm 10$ & $ 25 \pm 20 \pm 1$ \\
\hline
 $N_s(B^- \rightarrow D^{*0}~\ell^-~\bar{\nu})$   & $302 \pm 32 \pm 13$ &
($5\pm0.5$) \\
 $N_c(B^- \rightarrow D^{*0}~\ell^-~\bar{\nu})$   & ($14\pm14\pm7$) &
$20 \pm 20 \pm 1$ \\
\end{tabular}
\end{table}

\begin{table}
\caption{ \label{table:efficiency}
 Contributions to the fractional errors of the detection efficiency
for the $\bar{B}^0 \rightarrow D^{*+}~\ell^-~\bar{\nu}$ and
$B^- \rightarrow D^{*0}~\ell^-~\bar{\nu}$ decays. Errors common to both modes
are entered only once in the column labeled ``Both''.}
\begin{tabular}{cccc}
Source  & \multicolumn{3}{c}{ $\Delta\epsilon_s / \epsilon_s$ (\%) } \\
\cline{2-4}
                    & $\bar{B}^0 \rightarrow D^{*+}~\ell^-~\bar{\nu}$~~~~~~  &
Both~~~~~~  & $B^- \rightarrow D^{*0}~\ell^-~\bar{\nu}$~~~~~~  \\
\hline
Form Factors                        &               & $3.0$ &              \\
$\delta_m$ Sideband Normalization   & $0.8$         &       & $1.7$        \\
Variation of Cuts                   & $2.3$         &       & $1.0$        \\
Lepton Efficiency                   &               & $2.2$ &              \\
$D^0 \to K^- \pi^+$ Efficiency      &             & ~~$4.0~^*$ &           \\
Slow $\pi^+$ Efficiency             &               & $5.0$ &              \\
Slow $\pi^0/\pi^+$ Efficiency Ratio & --            &       & $7.0$        \\
\hline
Total Uncorrelated                  & $2.3$         &       & $7.3$       \\
Total Correlated                    &               & $7.4$ &             \\
\end{tabular}
$*$ This error is correlated with the systematic error of the CLEO II
$D^0 \to K^-\pi^+$ branching fraction.
\end{table}

\begin{table}
\caption{ \label{table:results} Results for product branching fractions.}
\begin{tabular}{lc}
Product Branching Fraction & Result $\times 10^4$ \\
\hline
${\cal B}(\Upsilon(4S) \rightarrow B^0\bar{B}^0)
{\cal B}(\bar{B}^0 \rightarrow D^{*+}~\ell^-~\bar{\nu}){\cal B}
(D^{*+}\rightarrow D^0\pi^+){\cal B}(D^0 \rightarrow K^-\pi^+)$
& $6.0 \pm0.43 \pm0.55$ \\
${\cal B}(\Upsilon(4S) \rightarrow B^+ B^-)
{\cal B}(\bar{B}^0 \rightarrow D^{*+}~\ell^-~\bar{\nu}){\cal B}
(D^{*0}\rightarrow D^0\pi^0){\cal B}(D^0 \rightarrow K^-\pi^+)$
& $6.4 \pm0.68 \pm0.73$ \\
\hline
${\cal B}(\bar{B} \rightarrow D^{*+}~\ell^-~\bar{\nu})
{\cal B}(D^{*+}\rightarrow D^0\pi^+){\cal B}(D^0 \rightarrow K^-\pi^+)$
& $1.6 \pm 0.7 \pm 0.3$ \\
${\cal B}(\bar{B} \rightarrow D^{*0}~X~\ell^-~\bar{\nu})
{\cal B}(D^{*0}\rightarrow D^0\pi^0){\cal B}(D^0 \rightarrow K^-\pi^+)$
& $1.4 \pm 1.5 \pm 0.3$
\end{tabular}
\end{table}

\begin{table}
\caption{ \label{table:br}
Comparison with previously published results for
${\cal B}(\bar{B}^0 \rightarrow D^{*+}~\ell^-~\bar{\nu})$ and
${\cal B}(B^- \rightarrow D^{*0}~\ell^-~\bar{\nu})$.
All previous results have been rescaled to use the CLEO II $D^0$ and
$D^*$ branching ratios (except for the CLEO1.5 $B^-$ number which depends
nonlinearly on these branching fractions), and use $f_{+-} = f_{00} = 0.5$.}
\begin{tabular}{lcc}
Experiment  &  ${\cal B}(\bar{B}^0 \rightarrow D^{*+}~\ell^-~\bar{\nu})$ (\%) &
 ${\cal B}(B^- \rightarrow D^{*0}~\ell^-~\bar{\nu})$ (\%) \\
\hline
CLEO II ($\tau_{B+}/\tau_{B^0} = 1.10 \pm 0.11$)  &
$4.57 \pm 0.29 \pm 0.41 \pm 0.22~^*$
                                      & $5.03 \pm 0.32 \pm 0.45 \pm 0.24~^*$ \\
CLEO II ($f_{+-} = f_{00} = 0.5$)          & $4.49 \pm 0.32 \pm 0.39$
                                           & $5.13 \pm 0.54 \pm 0.64$ \\
ARGUS \cite{oldargusIII,oldargusII}        & $4.9 \pm0.5 \pm 0.6$
                                           & $6.4 \pm1.5 \pm 1.4$\\
ARGUS \cite{arguspartial}                 & $4.5 \pm0.3 \pm 0.4$ & \\
CLEO 1.5 \cite{oldcleo,crawford}           & $4.1 \pm0.5 \pm 0.6$
                                           & $4.1 \pm0.8 \pm 0.9$ \\
\end{tabular}
$*$ When $\tau_{B^+}/\tau_{B^0} = 1.10$ is used to extract the CLEO II
branching fractions, all errors
for the two different exclusive modes are fully correlated.
\end{table}

\begin{table}
\caption{\label{table:modused}
Predictions for the form factor ratios and their
$q^2$ or $y$ dependence, and the
predictions and $q^2$ dependence for the common
form factor $A_1$. The pole forms have been abbreviated as:
$P_1 = 1- q^2/6.34^2$ and
$P_2 = 1- q^2/6.73^2$ where $q$ is in GeV.
${\cal E}(q^2)$ stands for $\exp[-0.03(q^2_{max}-q^2)]$.
For the form factor ratios of the Neubert
model~\protect{\cite{neubertmodel}}
$N(y) = 2.5/(y+1)$.}
\begin{center}
\begin{tabular}{lcccc}
         & ISGW~\cite{isgw}
         & BSW~\cite{bsw}
         & KS~\cite{ks}
         & Neubert~\cite{neubert}  \\ \hline
$V(q^2)/A_1(q^2)$
           & 1.27
           &$1.09P_2/P_1$
           &$1.00/P_1$
           &$N(y)[1.35-0.22(y-1)+0.09(y-1)^2]$  \\
$A_2(q^2)/A_1(q^2)$ & 1.14
           & $1.06$
           & $1.00/P_1$
           & $N(y)[0.79+0.15(y-1)-0.04(y-1)^2]$ \\
$A_1(q^2)$
           &$0.94{\cal E}(q^2)$
           &$0.65/P_2$
           &$0.70/P_1$
           &$0.86[2/(y+1)]^{0.6}$  \\
\end{tabular}
\end{center}
\end{table}

\begin{table}
\caption{\label{table:vcbmod}
Model dependent predictions of the
$\bar{B} \rightarrow D^*~\ell~\bar{\nu}$ partial
width, detection efficiencies for each model, and $|V_{cb}|$ values derived
from the measured partial width. An additional $3.5\%$ systematic error due to
$B$ lifetime measurements is common to all the $|V_{cb}|$ values given.}
\begin{center}
\begin{tabular}{lcccc}
         & ISGW~\cite{isgw}
         & BSW~\cite{bsw}
         &KS~\cite{ks}
         & Neubert~\cite{neubert}  \\ \hline
$\Gamma(\bar{B} \rightarrow D^*~\ell~\bar{\nu})|V_{cb}|^{-2} $
                      &24.6~ps$^{-1}$
                      &21.9~ps$^{-1}$
                      &25.8~ps$^{-1}$
                      &29.0~ps$^{-1}$  \\ \hline
$\epsilon_{{\rm model}}/\epsilon_{{\rm Neubert}}$
                      &1.00
                      &0.97
                      &0.98
                      & 1 \\
$|V_{cb}|\times 10^3$
                      &$34.8 \pm 1.1 \pm 1.6 $
                      &$37.5 \pm 1.2 \pm 1.7 $
                      &$34.4 \pm 1.1 \pm 1.5 $
                      &$32.2 \pm 1.0 \pm 1.4 $   \\
\end{tabular}
\end{center}
\end{table}

\begin{table}
\caption{\label{table:fitvcb}
Values for $|V_{cb}|{\cal F}(1)$, $a^2$ and $b$ determined
by the fits to the $d\Gamma/d\tilde{y}$ distributions. The first error is
statistical and the second systematic (including $B$ lifetimes). The values
for the individual modes, $\bar{B}^0 \rightarrow D^{*+}~\ell^-~\bar{\nu}$ and
$B^- \rightarrow D^{*0}~\ell^-~\bar{\nu}$, assume $f_{00}=f_{+-}=0.5$.
The results of the combined fit (bottom two rows) are
independent of this assumption.}
\begin{center}
\begin{tabular}{lccc}
 Mode     & $|V_{cb}|{\cal F}(1) \times 10^3$
          & $a^2$
          & $b$   \\ \hline
$\bar{B}^0 \rightarrow D^{*+}~\ell^-~\bar{\nu}$  &$34.7 \pm 2.5 \pm 1.8$
             &$0.80 \pm0.17 \pm 0.08$
             &0.0 \\
$B^- \rightarrow D^{*0}~\ell^-~\bar{\nu}$  &$35.7 \pm 2.8 \pm 2.4$
             &$0.94 \pm 0.20 \pm 0.08$
             &$0.0$   \\ \hline
$\bar{B} \rightarrow D^*~\ell~\bar{\nu}$
             &$35.1 \pm 1.9 \pm 1.9$
             &$0.84 \pm 0.13 \pm 0.08$
             &$0.0$   \\

             &$35.3 \pm 3.2 \pm 3.0 $
             &$0.92 \pm 0.64 \pm 0.40$
             &$0.15 \pm 1.24 \pm 0.90$   \\
\end{tabular}
\end{center}
\end{table}

\begin{table}
\caption{\label{table:sysvcb}
Estimates of the systematic error on $V_{cb}$ and $a^2$
determined from the fits to the $d\Gamma/dy$ distribution. The fractional
error on $V_{cb}$ is the same as the fractional error on the product
$|V_{cb}|{\cal F}(1)$
(the uncertainty in theoretical estimates of ${\cal F}(1)$
is not included in this table).
The error on the absolute efficiency is
taken from Table III, excluding the $4\%$ contribution from $K\pi$
reconstruction, which is also excluded from the error for ${\cal B}_{D^0}$. }
\begin{center}
\begin{tabular}{l|cc|cc|cc}
 source               &
\multicolumn{2}{c|}{$\bar{B}^0 \rightarrow D^{*+}~\ell^-~\bar{\nu}$}
		      &
\multicolumn{2}{c|}{$B^- \rightarrow D^{*0}~\ell^-~\bar{\nu}$}
		      &
\multicolumn{2}{c}{$\bar{B} \rightarrow D^*~\ell~\bar{\nu}$} \\

&$\Delta V_{cb}/ V_{cb} $&$ \Delta a^2/a^2$
&$\Delta V_{cb}/ V_{cb} $&$ \Delta a^2/a^2$
&$\Delta V_{cb}/ V_{cb} $&$ \Delta a^2/a^2$
\\ \hline
$B$ lifetime          & $2.9\%$ &      & $3.6\%$ &       &$2.3\%$ &  \\
Absolute Efficiency
                      & $3.3\%$ &      & $4.8\%$ &       & 3.3\%    &  \\

${\cal B}_{D^*} \times {\cal B}_{D^0}$
                      & $1.9\%$ &     &  $3.4\%$  &     & $2.1\%$  &    \\
Slow $\pi$ efficiency  shape
                      & $1.2\%$ & $3.5 \%$ &     &  & $0.8\%$  & $2.4\%$   \\
Fitting systematics
                      & $3\%$ & $9\%$       &  $3\%$ &  $9\%$
& $3\%$ & $9\%$    \\ \hline
Total
                      & $5.8\%$ & $9.7\%$    &  $7.5\%$ & $9.0\%$
& $5.5\%$ & $9.3\%$    \\
\end{tabular}
\end{center}
\end{table}

\begin{table}
\caption{\label{table:abmodel}
Model predictions for the form factor parameters $a^2$ and
$b$. These values were determined by fitting the $d\Gamma/dy$ distributions
predicted by each model. Results are given with and without $b$ fixed to zero.}
\begin{center}
\begin{tabular}{cccccc}
Parameter & This Exp.
          & ISGW~\cite{isgw}
          & BSW~\cite{bsw}
          & KS~\cite{ks}
          & Neubert~\cite{neubert}  \\ \hline
$a^2~(b = 0)$
       &$0.84\pm0.13\pm0.08$& $0.91$ & $0.77$ & $0.83$ & $0.48$ \\
\hline
$a^2$  &$0.92\pm 0.64 \pm 0.40$& $0.88$ & $0.84$ & $1.10$ & $0.59$ \\
$b$    &$0.15\pm 1.24 \pm 0.90$& $-0.1$ & $0.1$  & $0.6$  & $0.28$ \\
\end{tabular}
\end{center}
\end{table}

\begin{table}
\caption{\label{table:dgdqsquared}
Values for the differential decay rate at selected $q^2$ points.
These were calculated from
Eqs.\ (\protect{\ref{eq:neurate}}) and (\protect{\ref{eqn:three}}) with the
results of the maximum likelihood fit given in
Eqs.\ (\protect{\ref{eqn:vcbfbds}}) and (\protect{\ref{eqn:asquared}}).
The errors are combined statistical and systematic.}
\begin{center}
\begin{tabular}{ccccc}
   $q^2$                             & $m_{\pi^+}^2$   & $m_{\rho}^2$
                                     & $m_{a_1}^2$     & $m_{D_s}^2$ \\ \hline
$d\Gamma/dq^2$ (ns$^{-1}$GeV$^{-2}$) & $1.64\pm0.23$   & $1.90\pm0.25$
                                     & $2.32\pm0.31$   & $3.17\pm0.44$
\end{tabular}
\end{center}
\end{table}


\begin{thebibliography} {99}
\bibitem{ell}
We use the symbol $\ell$ and the term lepton to denote either an electron or a
muon.
\bibitem{chgconj}
Throughout this paper, charge conjugate decay modes are implied.
\bibitem{othermodes}
We investigated the feasibility of including other channels such as
$D^0 \rightarrow K^-\pi^+\pi^+\pi^-$ and $D^0 \rightarrow K^-\pi^+\pi^0$,
but the statistical significance of these modes is small
due to the increased combinatoric backgrounds. Additionally, inclusion
of these modes introduces new systematic errors, which offset the small
statistical gain. Thus the overall significance of our result would not
improve with more $D^0$ decay modes.
\bibitem{signalregion}
The Signal Region $(MM^2,C)$ boundaries are defined by the points $(0,0)$,
$(-1.5,1.4)$ and $(1.1,1.1)$.
\bibitem{badbarrel}
The calorimeter resolution quoted is for the region
$|\cos\theta_{\gamma}|<0.71$.
The resolution is slightly worse for $0.71<|\cos\theta_{\gamma}|<0.8$
\bibitem{hadronic}
To be classified as hadronic, an event must have at
least 3 charged tracks originating
near the interaction point, and the total visible
energy of the event must be greater
than 0.15 $E_{CM}$, where $E_{CM}$ is the center of mass energy.
\bibitem{delmfits}
The $\delta_m$ distributions are fit to
Gaussian signal functions and a phase space
background shape given by
$a_1(\delta_m-m_{\pi})^{1/2}+a_2(\delta_m-m_{\pi})^{3/2}$,
where $a_1$ and $a_2$ are free parameters and $m_{\pi}$ is the charged or
neutral pion mass. The signal widths from such fits are
$\sigma=1.01$~MeV and $\sigma=0.87$~MeV
for $D^{*+}$ and $D^{*0}$ respectively.
The sideband scale factors calculated from these fits are
$0.40 \pm0.03$ for $D^{*+}$ and $0.36 \pm0.01$ for $D^{*0}$.
\bibitem{beauty}
The tag lepton is required to have momentum above 1.8~GeV. This momentum cut
excludes leptons that were not directly produced by a $B$ meson decay.
\bibitem{neubertmodel}
By the ``Neubert model'' we mean the estimate of the shape of the form
factor ${\cal F}(y)$ from
Ref.\ \protect{\cite{neubert}}. This involves the use
of a theoretically-inspired form for the Isgur-Wise function with one free
parameter, $\rho^2 \approx 0.8$, which is estimated using QCD sum rules.
This is not regarded as a precision estimate.
Note that HQET makes no predictions about the shape of the Isgur-Wise function.
The expressions for the $A_1$, $A_2$ and $V$ form factors in this model have
been provided in Table~\protect{\ref{table:modused}}.
\bibitem{k3p}
Some previous measurements used the $D^0 \to K^-\pi^-\pi^+\pi^+$
decay channel as well as $D^0 \to K^-\pi^+$, but we have scaled all
results by the shift in the $D^0 \to K^-\pi^+$ branching fraction.
\bibitem{nomass}
An additional form factor is required for non zero lepton mass.
The $e$ and $\mu$ masses are neglected relative to the $B$ and $D^*$
masses.
\bibitem{etaa}
Neubert writes the form factor ${\cal F}(y)$ as $\eta_A\hat{\xi}(y)$,
where $\eta_A$ is a calculable radiative correction for hard gluon
processes ($\eta_A = 0.986\pm0.006$~\protect{\cite{neubert}})
and $\hat{\xi}(y)$ is the non perturbative part of the form factor.
\bibitem{slope}
In this parameterization, whenever $b=0$, $a^2$ is
equivalent to the slope, $\hat{\rho}^2$, as defined in Ref.\
\protect{\cite{neubert}} p. 163.
\bibitem{newcuts}
The event selection for the unbinned likelihood fits use $M_{K\pi}$ cuts
not present in background estimates listed in
Tables~\protect{\ref{table:numbers0}}
and \protect{\ref{table:numbers-}}. We therefore scale the numbers from
these tables according to the change in $\delta_m$
sideband subtracted raw yields caused by the $M_{K\pi}$ cuts.
This is an effect of order 1\%.
\bibitem{correlation}
The correlation coefficient from the linear fit is
$\rho_{|V_{cb}|,a^2} = 0.90$. For the quadratic fit the correlations
are: $\rho_{|V_{cb}|,a^2} = 0.92$, $\rho_{|V_{cb}|,b} = 0.84$,
$\rho_{a^2,b} = 0.98$.
\bibitem{liferror}
The result for $|V_{cb}|{\cal F}(1)$ from the
fit to the combined data scales as
$1/\sqrt{\tau_{B^+}+\tau_{B^0}}$, as long as $\tau_{B^+}/\tau_{B^0}$
remains constant.
A change in the lifetime ratio has a non
trivial effect because it modifies the
weights of the charged and neutral $B$ components in the
$d\Gamma/d{\tilde y}$ fit.
By performing different fits with varying lifetime ratios within $2\sigma$
of the central value~\protect{\cite{lifetimes}}
(keeping $\tau_{B^+}+\tau_{B^0}$ fixed)
we find, $\Delta |V_{cb}|{\cal F}(1) = 0.0012x - 0.003x^2$, where
$x = \tau_{B^+}/\tau_{B^0} - 1.1$.
\end{thebibliography}
\end{document}